\documentclass[lettersize,journal]{IEEEtran}
\usepackage{amsmath,amsfonts}
\usepackage{algorithmic}
\usepackage{algorithm}
\usepackage{array}
\usepackage[caption=false,font=normalsize,labelfont=sf,textfont=sf]{subfig}
\usepackage{textcomp}
\usepackage{xcolor}
\usepackage{pifont}
\usepackage{booktabs}
\usepackage{stfloats}
\usepackage{url}
\usepackage{verbatim}
\usepackage{graphicx}
\usepackage{multirow}
\usepackage{multicol}
\usepackage{cite}
\usepackage{tabularx}
\usepackage{longtable}
\definecolor{darkgreen}{rgb}{0.0, 0.5, 0.0}
\definecolor{darkred}{rgb}{0.5, 0.0, 0.0}
\definecolor{darkyellow}{HTML}{DAA520}
\newcommand{\cmark}{\textcolor{darkgreen}{\ding{51}}}      
\newcommand{\xmark}{\textcolor{darkred}{\ding{55}}}        
\newcommand{\tmark}{\textcolor{darkyellow}{$\triangle$}}   

\begin{document}

\title{Advances in Speech Separation: Techniques, Challenges, and Future Trends}
\author{
    Kai Li\textsuperscript{1, \textdagger, \textdaggerdbl},
    Guo Chen\textsuperscript{1, \textdaggerdbl},
    Wendi Sang\textsuperscript{2},
    Yi Luo\textsuperscript{3},
    Zhuo Chen\textsuperscript{4},
    Shuai Wang\textsuperscript{5},
    Shulin He\textsuperscript{6},
    Zhong-Qiu Wang\textsuperscript{6},
    Andong Li\textsuperscript{7},
    Zhiyong Wu\textsuperscript{8},
    and Xiaolin Hu\textsuperscript{1, *}
\thanks{Manuscript received April 19, 2021; revised August 16, 2021.}
\thanks{This work was supported by the National Key Research and Development Program of China under Grant 2021ZD0200301 and the National Natural Science Foundation of China under Grant U2341228.}
\thanks{\textsuperscript{1}Kai Li, Guo Chen, and Xiaolin Hu are with the Department of Computer Science and Technology, Tsinghua University, Beijing, China.}
\thanks{\textsuperscript{2}Wendi Sang is with the School of Computer Technology and Application, Qinghai University, Xining, China.}
\thanks{\textsuperscript{3}Yi Luo is an independent author, Shenzhen, China (e-mail: y.luo@columbia.edu).} 
\thanks{\textsuperscript{4}Zhuo Chen is with ByteDance.}
\thanks{\textsuperscript{5}Shuai Wang is with Nanjing University, Suzhou, China.}
\thanks{\textsuperscript{6}Shulin He and Zhong-Qiu Wang are with the Southern University of Science and Technology, Shenzhen, China.}
\thanks{\textsuperscript{7}Andong Li is with the Institute of Acoustics, Chinese Academy of Sciences, Beijing, China.}
\thanks{\textsuperscript{8}Zhiyong Wu is with the Shenzhen International Graduate School, Tsinghua University, Shenzhen, China.}
\thanks{\textsuperscript{\textdagger}Project leader.}
\thanks{\textsuperscript{\textdaggerdbl}Equal contribution. (Kai Li and Guo Chen contributed equally to this work.)} 
\thanks{\textsuperscript{*}Corresponding author: Xiaolin Hu (e-mail: xlhu@tsinghua.edu.cn).} 
}

\markboth{Journal of \LaTeX\ Class Files,~Vol.~14, No.~8, August~2021}%
{Shell \MakeLowercase{\textit{et al.}}: A Sample Article Using IEEEtran.cls for IEEE Journals}


\maketitle

\begin{abstract}
The field of speech separation, targeting the challenging ``cocktail party problem'', has witnessed revolutionary advances along with the development of deep neural networks (DNNs). Speech separation can be employed in standalone applications, enhancing speech clarity in complex acoustic environments. Additionally, it can function as a crucial pre-processing method for other speech processing tasks, such as speech recognition and speaker recognition. Despite numerous publications addressing this challenge, current literature surveys and reviews tend to focus narrowly on specific architectural designs or isolated learning approaches, creating a fragmented understanding of this rapidly evolving domain. This fragmentation underscores the urgent need for a unified, comprehensive survey that captures the field's breadth and recent innovations. This survey addresses this critical gap by providing a systematic and holistic examination of DNN-based speech separation techniques. Our work differentiates itself from existing surveys and reviews: (I) \textit{Comprehensive perspective}: Instead of only cataloging architectural variations, we systematically investigated higher-level learning paradigms, comprehensively covering separation scenarios with known or unknown number of speakers, comparative analysis of supervised, self-supervised, and unsupervised learning frameworks, and detailed examination of various architectural components from front-end encoders to back-end estimation strategies. (II) \textit{Rigorous timeliness}: By covering the cutting-edge developments from the most recent years, we provide up-to-date coverage to ensure researchers have access to the most current methodological innovations and performance benchmarks. (III) \textit{Unique insights}: Beyond mere summarization, we critically evaluate technological trajectories, identify emerging research patterns, and highlight promising directions including domain-robust frameworks, computationally efficient architectures, multimodal integration approaches, and novel self-supervised paradigms. (IV) \textit{Fair quantitative evaluation}: We provide fair quantitative evaluations on standard datasets, aiming to provide clear and reliable performance benchmarks for the research community, thereby revealing the true capabilities and limitations of different methods. This comprehensive survey aims to serve as an accessible reference for both experienced researchers and newcomers navigating the complex landscape of speech separation.
\end{abstract}


\section{Introduction}
\label{sec:intro}
\IEEEPARstart{I}{n} the human auditory system, the “cocktail party effect” exemplifies our remarkable ability to selectively attend to a target speaker while suppressing interfering sources in noisy environments \cite{cherry1953some,arons1992review,bregman1990auditory}. However, endowing machines with a comparable level of separation capability remains one of the most challenging topics in the field of speech processing.

\begin{figure*}[!t]
\centering
\includegraphics[width=1.0\linewidth]{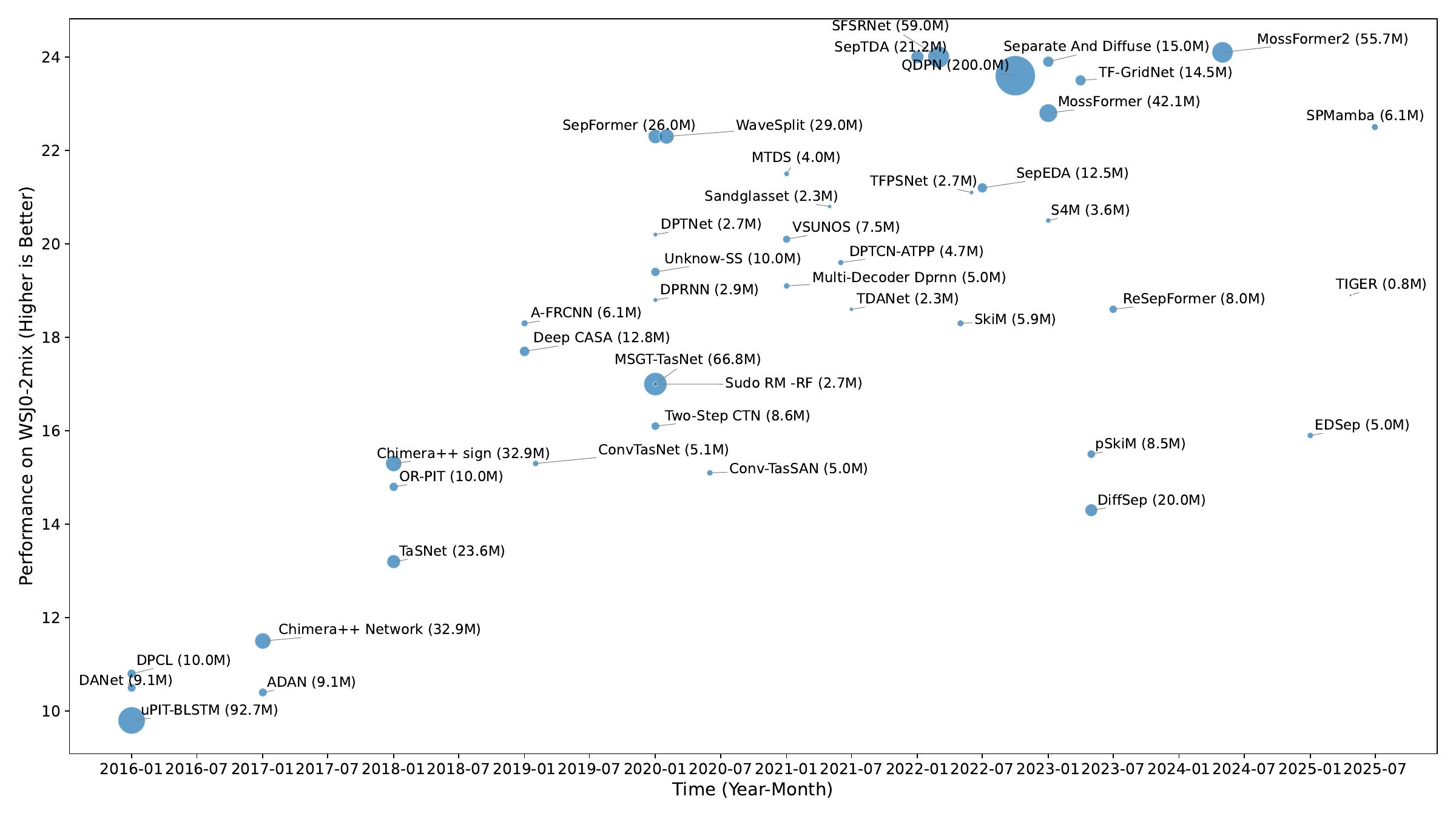}
\caption{Speech Separation Model Performance on WSJ0-2mix over Time. The size of the points represents the number of parameters of the model.}
\label{fig:history}
\vspace{-15pt}
\end{figure*}

Over the past decades, conventional approaches have primarily relied on statistical models and signal processing techniques, such as Independent Component Analysis (ICA) \cite{hyvarinen2000independent}, Non-negative Matrix Factorization (NMF) \cite{cichocki2006new}, and heuristic spatial filtering methods \cite{tesch2023multi,mandel2017multichannel}. Although these methods perform well in certain scenarios, their separation performance and generalization ability are severely limited in real-world settings characterized by high nonlinearity, complex environments, and dynamically varying numbers of speech sources \cite{chen2017deep}. In recent years, a data-driven paradigm—spearheaded by deep neural networks (DNNs)—has triggered a significant shift, greatly advancing the state-of-the-art in speech separation systems. A host of architectures and methodologies have been proposed, encompassing Convolutional Neural Networks (CNNs) \cite{luo2019conv,tzinis2020sudo,hu2021speech}, Recurrent Neural Networks (RNNs) \cite{yu2017permutation,luo2020dual,chen2020dual,wang2023tf}, and more recently, variants based on self-attention mechanisms such as Transformers \cite{subakan2021attention,zhao2023mossformer,zhao2024mossformer2}, as well as generative models like Generative Adversarial Networks (GANs) \cite{goodfellow2020generative,li2021generative} and diffusion models \cite{fu2019metricgan,chen2023sepdiff,lutati2023separate}. These modern DNN-based approaches automatically learn deep hierarchical features and, through direct exposure to large-scale mixtures, substantially boost the performance of speech separation tasks. As visually illustrated in Figure~\ref{fig:history}, performance curves on the canonical WSJ0-2mix dataset highlight the remarkable progress driven by the evolution of architectures and algorithms \cite{hershey2016deep}.

Nevertheless, alongside rapid technological advances, application demands for speech separation have become increasingly diverse and complex. For example, meeting transcription and real-time communication require extremely low latency and online output; assistive hearing devices and mobile applications demand lightweight, efficient models with constrained resources. Meanwhile, real-world scenarios present persistent challenges such as an unknown and dynamically varying number of speakers, frequent speaker switching, as well as non-stationary noise and reverberation—all of which remain significant bottlenecks impeding the large-scale deployment of speech separation algorithms \cite{takahashi2019recursive,nachmani2020voice,zhu2021multi,chetupalli2022speech}.

\begin{table*}[!t]
\centering
\caption{A comparative analysis of recent surveys and reviews on deep learning-based speech separation. \cmark: Comprehensive coverage, \tmark: Partial coverage, \xmark: Limited or no coverage.}
\begin{tabular}{cccccccc}
\toprule
\textbf{Surveys and reviews (Year)}          & \textbf{Deep Learning Methods} & \textbf{Architecture} & \textbf{Topics} & \textbf{Evaluation} & \textbf{Datasets} & \textbf{Platforms} & \textbf{Results} \\ 
\midrule
\cite{qian2018past} (2018) & \tmark & \tmark & \xmark & \xmark & \xmark & \xmark & \xmark \\
\cite{wang2018supervised} (2018)   & \tmark & \tmark & \xmark & \tmark & \tmark & \xmark & \xmark \\
\cite{mirbeygi2022speech} (2022)   & \tmark & \tmark & \xmark & \tmark & \tmark & \xmark & \xmark \\
\cite{agrawal2023review} (2023)    & \tmark & \tmark & \xmark & \tmark & \tmark & \tmark & \xmark \\
\cite{ochieng2023deep} (2023)      & \cmark & \cmark & \tmark & \tmark & \tmark & \xmark & \xmark \\
\cite{ansari2023survey} (2023)     & \cmark & \tmark & \xmark & \tmark & \tmark & \xmark & \tmark \\
Ours (2025)                               & \cmark & \cmark & \cmark & \cmark & \cmark & \cmark & \cmark \\
\bottomrule
\end{tabular}
\label{tab:surveys}
\end{table*}

Given the rapid evolution of this field, a comprehensive and systematic survey of deep learning-based speech separation is not just necessary but urgently needed. This paper aims to provide an authoritative guide for researchers and practitioners, clarifying the current research landscape and critically assessing the strengths, limitations, and future trajectories of cutting-edge technologies in addressing real-world challenges.

While existing surveys and reviews \cite{qian2018past,wang2018supervised,agrawal2023review,mirbeygi2022speech,ochieng2023deep,ansari2023survey} offer valuable summaries, they struggle to keep pace with the relentless innovation in this domain. As illustrated in Table~\ref{tab:surveys}, the literature still lacks a single, up-to-date survey that systematically integrates the latest complex model architectures, emerging modeling paradigms (e.g., unsupervised and self-supervised learning), standardized evaluation frameworks, and critical dataset platforms.

Specifically, early seminal surveys and reviews like Qian et al.~\cite{qian2018past} and Wang and Chen~\cite{wang2018supervised} established a solid foundation by explaining the cocktail party problem and initial deep learning approaches~\cite{hershey2016deep,yu2017permutation}. However, their historical context is also their primary limitation. They entirely miss the disruptive innovations of the last five years, such as advanced end-to-end models, unsupervised paradigms, and pre-trained models. Furthermore, they severely lack comprehensive discussions on practical aspects like evaluation metrics (e.g., subjective metrics beyond SDR~\cite{le2019sdr} and PESQ~\cite{rix2001perceptual}), datasets, and comparative performance analysis, rendering their utility limited for contemporary research.

More critically, even the most recent surveys and reviews and reviews by Agrawal et al.~\cite{agrawal2023review} and Ansari et al.~\cite{ansari2023survey} suffer from a fundamental flaw. While broader in scope, their approach to results analysis is often a simple aggregation of performance metrics reported in original papers. Due to disparate experimental setups, training data, and evaluation scripts, these metrics, gathered from non-standardized environments, are not directly comparable and can even be misleading. This creates significant confusion for researchers seeking objective assessments and technology selection. Therefore, establishing a fair comparison framework under a unified standard is an urgent, unmet need in this field.

Finally, the existing literature has also largely overlooked in-depth discussions on several topics crucial for real-world applications: the trade-off between model compactness and real-time deployment, the inherent challenges of different learning paradigms, and a critical analysis of scenario coverage and biases in existing datasets. To address all the aforementioned challenges and fill these gaps, this work presents a comprehensive and systematic survey of deep neural network-based speech separation technologies. We provide an in-depth analysis that traces the evolution of the field from fundamental theories to cutting-edge architectures, with particular emphasis on the core designs and technical strengths of mainstream models such as dual-path networks~\cite{luo2020dual}, U-Net variants~\cite{tzinis2020sudo,xu2024tiger}, and Transformer-based architectures~\cite{subakan2021attention,zhao2024mossformer2}.

Distinct from prior surveys and reviews, our primary contribution lies in the implementation of rigorous and reproducible benchmarking of representative models within a unified experimental framework. By conducting fair quantitative evaluations on both standard datasets (e.g., LibriMix~\cite{cosentino2020librimix}, WHAM!~\cite{wichern2019wham}, REAL-M~\cite{subakan2022real}) and newly introduced challenging datasets (e.g., LRS2-2Mix~\cite{li2024audio}, SonicSet~\cite{li2024sonicsim}), we aim to provide the research community with a clear and reliable performance benchmark\footnote{\url{https://cslikai.cn/Speech-Separation-Paper-Tutorial/}}, thereby revealing the true capabilities and limitations of different methodological approaches. In addition, this survey systematically summarizes the functionalities of open-source toolkits such as Asteroid~\cite{pariente2020asteroid}, SpeechBrain~\cite{speechbrain}, and WeSep~\cite{wang24fa_interspeech}, and discusses key challenges and future directions. We anticipate that this work will offer a comprehensive knowledge map serving as both an introductory guide for new researchers and a decision-making reference for experienced engineers in applications such as meeting assistants, wearable hearing devices, and augmented reality.

\section{Problem Formulation}
\label{sec:formulation}
\subsection{The Cocktail Party Problem}
\label{sec:cocktail-party}

The cocktail party effect refers to the human ability to selectively focus on a single auditory source, such as a particular voice or sound, amidst competing background noise \cite{cherry1953some,haykin2005cocktail}. This capacity is critical for navigating real-world auditory environments, where sound sources often overlap in both time and space. The phenomenon is underpinned by the brain’s ability to suppress irrelevant auditory inputs while isolating and prioritizing meaningful acoustic signals—a fundamental process in auditory scene analysis \cite{arons1992review,mcdermott2009cocktail}. At its core, the cocktail party effect exemplifies the interaction between bottom-up sensory processing and top-down attentional modulation, mediated by the dynamic interplay within hierarchical brain networks \cite{bregman1990auditory,carlyon2004brain,cusack2004effects,carlyon2001effects,kuo2022inferring}.

The brain integrates temporally synchronized auditory features to form coherent perceptual objects \cite{elhilali2009temporal}. Neuronal populations encoding sound source attributes—like pitch, location, and harmonics—show phase-locked oscillations, binding these features into a unified stream while segregating asynchronous distractors \cite{divenyi2004speech}. This involves feedforward encoding of features in primary auditory cortex (A1) \cite{li2014feedforward} and feedback from higher regions (e.g., prefrontal and parietal cortices) that modulate integration based on attention \cite{power2012time}. For example, focusing on a speaker enhances neural gain for their voice traits \cite{houde2011speech}, amplifying its representation amid noise and enabling target speech tracking.

Top-down attentional control enables selective auditory processing \cite{awh2012top,mai2025linear}. The prefrontal cortex (PFC) and anterior cingulate cortex (ACC) produce modulatory signals that bias competition among auditory streams, favoring behaviorally relevant ones \cite{milham2003competition}. This bias manifests as enhanced phase alignment of theta (4–8 Hz) and gamma (30–80 Hz) oscillations in auditory cortex, synchronizing neural activity with attended speech rhythms \cite{gourevitch2020oscillations}. For instance, speech's syllabic rhythm (3–8 Hz) entrains theta oscillations in the superior temporal gyrus (STG), creating a temporal template that tracks the target speaker's acoustic envelope \cite{patel2022interaction}. Concurrently, gamma oscillations bind fine spectral-temporal features (e.g., formants, phonemes) into a unified percept \cite{patel2022interaction,li2024does}. Notably, attention amplifies target voice representations while suppressing unattended stimuli, as seen in attenuated N1 event-related potentials for ignored streams, reducing neural responsiveness below baseline.

Hierarchical processing in the auditory system refines selective attention \cite{shinn2008selective,lakatos2013spectrotemporal,sussman2017auditory}. Early stages in the brainstem and midbrain \cite{mesgarani2012selective,dahmen2010adaptation,bronkhorst2015cocktail,du2023decoding}, especially the superior olivary complex, use binaural cues—such as interaural time differences (ITDs) and interaural level differences (ILDs)—to spatially segregate sound sources. These pre-attentive processes boost the signal-to-noise ratio (SNR) of targets via spatial unmasking. In the auditory cortex, processing advances from spectrotemporal analysis in A1 to advanced feature extraction in non-primary areas like the planum temporale (PT) \cite{hamilton2020topography}. The PT integrates spatial and spectral data to separate overlapping streams and shows enhanced sensitivity to auditory edges—sudden acoustic energy shifts signaling new sources—enabling the parsing of complex scenes into distinct perceptual streams.


The cocktail party effect arises from a cascade of neural operations: temporal coherence binds acoustic features into perceptual streams, attentional gain modulates their cortical representation, hierarchical auditory processing separates spatial and spectrotemporal cues, and large-scale networks integrate these processes into a coherent perceptual experience. These mechanisms collectively enable the brain to resolve the ambiguity of complex auditory scenes, prioritizing meaningful signals while filtering out noise—an ability that remains unmatched by even the most advanced computational models.

\begin{figure*}[!t]
\centering
\includegraphics[width=0.9\linewidth]{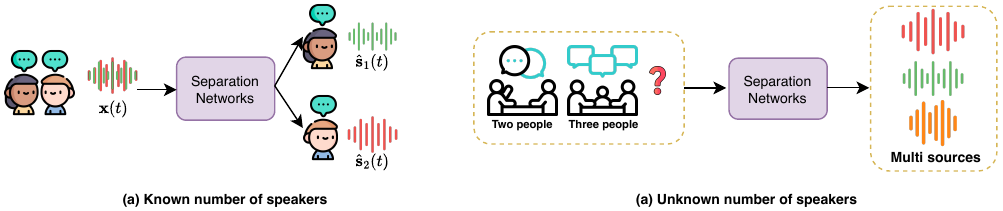}
\caption{Overview of speech separation with known/unknown source counts.}
\label{fig:known-unknown}
\end{figure*}

\subsection{Speech Separation}

As stated in the previous subsection, humans possess a natural ability to distinguish different sound sources in noisy environments, a phenomenon known as the \textit{cocktail party effect} \cite{cherry1953some}. In speech signal processing, this corresponds to the task of \textit{speech separation} \cite{wang2018supervised}. The objective of this task is to separate individual speech source signals $\{\mathbf{s}_i \in \mathbb{R}^{1 \times T} \mid i \in [1, C]\}$ from a mixed audio input $\mathbf{x} \in \mathbb{R}^{1 \times T}$, where $T$ denotes the length of the audio signal, and $C$ represents the number of speakers in the mixture. The mixed signal typically contains multiple sound sources that overlap substantially in both time and frequency domains, thereby significantly increasing the complexity of the separation process. Formally, the task can be represented as:
\begin{equation}
    \mathbf{x} = \sum_{i=1}^{C}\mathbf{s}_i + \mathbf{n}
\end{equation}
where $\mathbf{n} \in \mathbb{R}^{1 \times T}$ denotes background noise.

The input of the speech separation task can be represented in the time domain or frequency domain. In the time-domain representation, the raw audio waveform is processed directly, usually using one-dimensional convolutional layers to extract high-dimensional embedding features \cite{luo2018tasnet}. In the frequency-domain representation, the signal is transformed into a time-frequency representation via short-time Fourier transform (STFT) \cite{hershey2016deep}. Regardless of the input representation used, current mainstream speech separation methods mostly adopt a modular processing pipeline, typically consisting of an encoder, separator, audio estimation, and decoder. The encoder first maps the input signal to a high-dimensional feature space; the separator decouples and distinguishes information from different sound sources in this space; then, through audio estimation (e.g., generating masks or directly estimating source signal features), representations of each source are obtained; finally, the decoder is responsible for reconstructing these representations into independent time-domain audio signals. This architecture provides a foundation for tackling subsequent challenges. For example, the input of speech separation methods can be single-channel or multi-channel audio data. In the single-channel scenario, the separation task is particularly challenging due to the lack of spatial information that helps differentiate between different sound sources. In contrast, multi-channel settings utilize microphone arrays to exploit spatial features, thereby improving separation performance. However, this setting also brings challenges related to hardware deployment and increased algorithmic complexity. 

Additionally, speech separation systems need to generate independent audio signals for each source and be capable of handling both known and unknown numbers of sources. This requirement is especially important in applications such as real-time voice communication \cite{divenyi2004speech}, conference systems \cite{rao2021conferencingspeech}, and human-computer interaction \cite{panda2017automated}, where systems must demonstrate adaptability in dynamic environments. Therefore, speech separation methods must strike a balance between signal quality, computational efficiency, and model generalization ability to meet the demands of diverse application scenarios.

\subsubsection{Speech Separation with Known Number of Sound Sources}

In the speech separation with the known number of sound sources, the objective is to separate $ C $ independent speech signals from a mixed audio input, as shown in Figure \ref{fig:known-unknown}, where $ C $ represents the known number of speakers. Specifically, this task involves learning a mapping $ f_\theta(\mathbf{x}) \to \{\hat{\mathbf{s}}_1, \hat{\mathbf{s}}_2, \dots, \hat{\mathbf{s}}_C\} $ via a separation network $ f_\theta(\cdot) $, such that the separated signals $ \hat{\mathbf{s}}_i $ closely approximate the corresponding ground-truth source signals $ \mathbf{s}_i $. In deep learning-based approaches, criterion-driven loss functions are often employed to enhance separation performance \cite{le2019sdr}. For instance, the Scale-Invariant Signal-to-Distortion Ratio (SI-SDR) loss function \cite{le2019sdr} optimizes at the signal level to improve the perceptual quality of the separated signals. The design of separation networks typically utilizes deep neural network architectures (see Section \ref{sec:system}), such as temporal convolutional networks (TCNs) \cite{lea2016temporal}, recurrent neural networks (RNNs) \cite{yu2019review} and Transformer-based models \cite{vaswani2017attention}. These architectures effectively capture long-term dependencies by modeling contextual information.

In this task, the output layer of the network is constrained to produce a fixed number of output channels $ C $. To address the speaker permutation ambiguity problem, strategies such as deep clustering \cite{hershey2016deep} or Permutation-Invariant Training (PIT) \cite{yu2017permutation} (detailed in Section \ref{sec:learning}) are commonly employed. These strategies enable accurate modeling and, when combined with advanced neural network architectures and optimization techniques, significantly improve the separation performance.

\subsubsection{Speech Separation with Unknown Number of Sound Sources}


The task of single-channel speech separation with an unknown number of speakers aims to recover $C$ independent speaker signals $\{\mathbf{s}_i\}_{i=1}^C$ from a mixture $\mathbf{x}$, as shown in Figure \ref{fig:known-unknown}. Unlike conventional separation tasks with fixed source counts, the core challenge lies in the joint optimization of dynamic output dimensions and separation termination conditions. In practical scenarios (e.g., meeting transcription, multi-party dialogue), the unknown and time-varying nature of $C$ renders traditional fixed-output-channel methods ineffective. Specific challenges include \cite{takahashi2019recursive}: 1) \textit{Exponential expansion of permutation ambiguity}: When handling variable outputs, conventional PIT must extend to asymmetric output configurations. 2) \textit{Trade-off between separation quality and termination robustness}: The separation process requires the adaptive generation of $C$-matched channels while avoiding under-separation (incomplete source extraction) or over-separation (redundant noise channels).

Existing approaches primarily address these challenges through recursive separation frameworks \cite{takahashi2019recursive,nachmani2020voice} and dynamic network architectures \cite{chazan2021single}. The former iteratively extracts individual speakers while updating residual signals (e.g., OR-PIT \cite{takahashi2019recursive}), formalized as:
\begin{equation}
    \mathbf{r}_k = f_{\theta}(\mathbf{r}_{k-1}), \quad \mathbf{r}_0 = \mathbf{x}
\end{equation}
where $f_{\theta}$ denotes the single-step separation network and $\mathbf{r}_k$ represents the residual signal at the $k$-th iteration. This approach dynamically terminates separation through threshold detection (e.g., energy ratio or confidence score), demonstrating generalization capability to unseen $C$ values (e.g., $C=4$) during testing. The latter category employs dynamic network architectures for flexible output generation. Representative methods include multi-decoder mixture-of-experts systems, where a shared encoder extracts mixed features before multiple expert decoders generate different output counts, with a speaker quantity classifier selecting optimal results  \cite{zhu2021multi}. Another line of work adapts speaker diarization concepts through dynamic attractor computation, where LSTMs generate candidate attractors followed by similarity thresholding for effective source count estimation \cite{chetupalli2022speech,lee2024boosting}. Current methods still face challenges in computational efficiency (e.g., error accumulation in recursive separation) and noise robustness (e.g., attractor selection sensitivity to reverberation). 

\section{Learning Paradigms}
\label{sec:learning}
\begin{table*}[!t]
\centering
\caption{A summary of speech separation methods based on different learning approaches. Methods are sorted in chronological order. For detailed content, refer to Section \ref{sec:learning}.}
\begin{tabular}{cc|m{13cm}}
\toprule
\multicolumn{2}{c|}{\textbf{Learning Approaches}}                      & \textbf{Methods}                                                                    \\ \midrule
\multicolumn{2}{c|}{Unsupervised Learning}                             & MixIT \cite{wisdom2020unsupervised}, VAE \cite{neri2021unsupervised}, TS-MixIT \cite{zhang2021teacher}, UNSSOR \cite{wang2023unssor}                            \\ \midrule
\multicolumn{1}{c|}{\multirow{2}{*}{Supervised Learning}} & Clustering & DPCL \cite{hershey2016deep}, DANet \cite{chen2017deep}, ADAN \cite{luo2017speaker}, Chimera++ Network \cite{wang2018alternative}, Chimera++ sign \cite{wang2019deep}, WaveSplit \cite{zeghidour2020wavesplit}          \\ 
\cmidrule{2-3}
\multicolumn{1}{c|}{}                                     & PIT        & PIT \cite{yu2017permutation}, uPIT-BLSTM \cite{kolbaek2017multitalker}, TaSNet \cite{luo2018tasnet}, Wave-UNet \cite{stoller2018wave}, SSGAN \cite{subakan2018generative}, SSGAN-PIT \cite{chen2018permutation}, CBLDNN-GAT \cite{li2018cbldnn}, Conv-TasNet \cite{luo2019conv}, Deep CASA \cite{liu2019divide}, OR-PIT \cite{takahashi2019recursive}, FurcaNeXt \cite{zhang2020furcanext}, VSUNOS \cite{nachmani2020voice}, DPRNN \cite{luo2020dual}, DPTNet \cite{chen2020dual}, Conv-TasSAN \cite{deng2020conv}, Two-Step TCN \cite{tzinis2020two}, SudoRM-RF \cite{tzinis2020sudo}, Multi-Decoder Dprnn \cite{zhu2021multi}, DPTCN-ATPP \cite{zhu2021dptcn}, MSGT-TasNet \cite{zhao2021multi}, TS-MixIT \cite{zhang2021teacher}, SepFormer \cite{subakan2021attention}, A-FRCNN \cite{hu2021speech}, Sandglasset \cite{lam2021sandglasset}, CDGAN \cite{li2021generative}, Unknow-SS \cite{chazan2021single}, QDPN \cite{rixen2022qdpn}, SFSRNet \cite{rixen2022sfsrnet}, MTDS \cite{qian2022efficient}, TDANet \cite{li2022efficient}, SkiM \cite{li2022skim}, TFPSNet \cite{yang2022tfpsnet}, SSL-SS \cite{huang2022investigating}, SepEDA \cite{chetupalli2022speech}, MossFormer \cite{zhao2023mossformer}, SepDiff \cite{chen2023sepdiff}, DiffSep \cite{scheibler2023diffusion}, Separate And Diffuse \cite{lutati2023separate}, PGSS \cite{li2023pgss}, pSkiM \cite{li2023predictive}, Diff-Refiner \cite{hirano2023diffusion}, CycleGAN-SS \cite{joseph2023cycle}, HuBERT \cite{fazel2023cocktail}, S4M \cite{chen2023neural}, TF-GridNet \cite{wang2023tf}, MossFormer2 \cite{zhao2024mossformer2}, TCodecSS \cite{yip2024towards}, CodecSS \cite{yip2024speech}, DIP \cite{wang2024speech}, Fast-GeCo \cite{wang2024noise}, SepTDA \cite{lee2024boosting}, Conv-TasNet GAN \cite{lakandri2024exploring}, ReSepFormer \cite{della2024resource}, EDSep \cite{dong2025edsep}, TIGER \cite{xu2024tiger}, SPMamba \cite{li2024spmamba} \\ \bottomrule
\end{tabular}
\label{tab:learning}
\end{table*}

\subsection{Unsupervised Learning}

Speech separation aims to extract individual source signals from a mixed recording and constitutes a long-standing challenge in the field of audio signal processing. Early explorations in this domain primarily pursued \textit{unsupervised} solutions, seeking to uncover latent sources in the absence of accessible ground-truth signals. As a pioneering approach, Independent Component Analysis (ICA) leverages the statistical independence among source signals to perform blind separation~\cite{hyvarinen2000independent}. Despite its foundational role, ICA demonstrates limited effectiveness in highly underdetermined, single-channel scenarios, and its performance relies on strong statistical assumptions that may not hold in real acoustic environments. Subsequent extensions such as Independent Vector Analysis (IVA) model dependencies among source vectors to enhance the processing of multi-channel signals, yet these methods still inherit dependence on statistical priors and require multi-microphone input~\cite{kitamura2016determined}. In parallel, matrix factorization-based methods, such as Non-negative Matrix Factorization (NMF), have emerged as alternatives~\cite{cichocki2006new}. NMF operates by decomposing the magnitude spectrum of the mixed signal into a set of basis vectors (i.e., spectral templates) and the corresponding activations. A core limitation of NMF is its restricted model expressiveness; the learned bases are typically static and struggle to capture the complex temporal variations present in speech, which impairs separation quality.

With the advent of deep learning, unsupervised methods have been revitalized via more powerful generative models. For instance, frameworks based on Variational Autoencoders (VAEs) achieve separation by learning probabilistic priors over speech signals within a latent space~\cite{neri2021unsupervised}. Recently, the ``mixture of mixtures" (MixIT) paradigm has made notable advances by training on mixtures of mixtures without access to source references~\cite{wisdom2020unsupervised}. However, MixIT suffers from over-separation issues and requires single-speaker data for convergence. To address this, semi-supervised extensions such as the teacher-student paradigm incorporate limited supervision~\cite{zhang2021teacher}, albeit still relying on a portion of annotated data. UNSSOR exploits multi-channel signals in over-determined settings to achieve unsupervised separation~\cite{wang2023unssor}, optimizing filters through convolutional prediction, but generalizing to underdetermined conditions remains challenging.

Although unsupervised approaches—by leveraging statistical priors or consistency objectives—offer elegant and data-efficient solutions, they are fundamentally constrained by the strength of their underlying assumptions. Consequently, their performance in complex acoustic scenarios often fails to meet the fidelity requirements of high-quality separation. These inherent limitations have driven the research community towards a significant paradigm shift to supervised learning, wherein models are directly trained on large-scale datasets of paired mixtures and their corresponding source signals.

\subsection{Supervise Learning}

\begin{figure}[!t]
\centering
\includegraphics[width=1.0\linewidth]{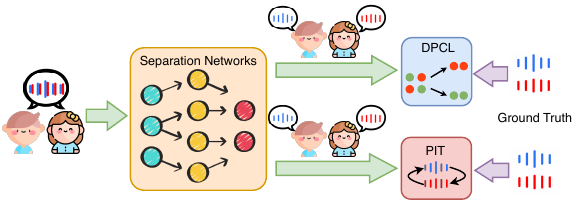}
\caption{The supervised speech separation workflow. The diagram presents two mainstream approaches: Deep Clustering (DPCL) and Permutation Invariant Training (PIT). The number of speakers is assumed to be 2.}
\label{fig:supervise}
\end{figure}

Supervised speech separation constitutes a dominant approach in the domain of audio signal processing, aiming to extract independent source signals from mixed acoustic inputs through model training, as shown in Fig.~\ref{fig:supervise}. These methods rely on a parallel dataset composed of mixed signals and their corresponding clean signals, enabling the model to learn a mapping function from the mixed signals to the constituent sources. One of the main challenges in this domain is the correct assignment of the components within a mixed speech signal to their corresponding source signals. This challenge arises from the inherent ambiguity in the order of speech components within mixed speech, leading to what is known as the label permutation problem \cite{hershey2016deep,yu2017permutation}. Taking the multi-speaker scenario as an example: due to the arbitrary nature of speaker order within mixed speech, there is no predetermined correspondence between the output channels of the separation network and the target speakers.

Current approaches to address this challenge primarily fall into two categories:
\begin{itemize}
    \item[1.] Deep clustering (DC) \cite{hershey2016deep}, which transform the separation problem into a high-dimensional embedding space, allowing time-frequency units belonging to the same source to cluster together;
    \item[2.] Permutation Invariant Training (PIT) \cite{yu2017permutation}, which evaluates all possible output-target permutations during training and selects the permutation scheme that minimizes the loss function.
\end{itemize}
These methods tackle the label permutation problem in supervised speech separation from different perspectives.

\subsubsection{Clustering Methods}

\begin{figure}[!t]
\centering
\includegraphics[width=1.0\linewidth]{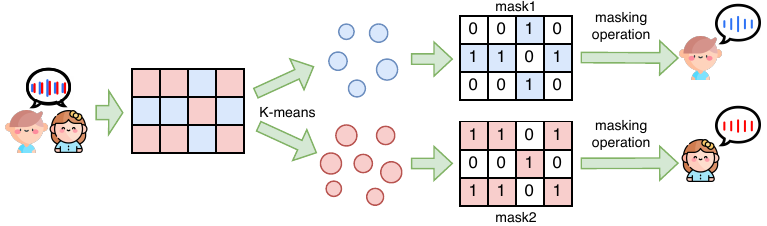}
\caption{The pipeline of the deep clustering method.}
\label{fig:dpcl}
\end{figure}

Unlike conventional clustering algorithms which do not require teaching signals, these deep clustering methods are supervised learning methods. Deep Clustering \cite{hershey2016deep} circumvents this problem through an innovative approach. Instead of directly predicting separated signals or masks, it trains a deep neural network to map each time-frequency (T-F) unit (indexed by $i=(t,f)$) in the mixed speech spectrogram (typically obtained through Short-Time Fourier Transform, STFT) to a high-dimensional embedding space, generating embedding vectors $\mathbf{v}_i \in \mathbb{R}^D$ \cite{hershey2016deep}. The core idea is to ensure that embedding vectors from the same source (e.g., the same speaker) are positioned close to each other in the embedding space, while those from different sources remain distant from each other, as shown in Fig.~\ref{fig:dpcl}. To achieve this, Deep Clustering typically optimizes a loss function based on the difference between an affinity matrix constructed from embedding vectors and an ideal affinity matrix. Specifically, all T-F unit embedding vectors ($N=TF$) are stacked into a matrix $\mathbf{V} \in \mathbb{R}^{N \times D}$, and the corresponding ideal source assignments (e.g., binary labels based on which source dominates in that T-F unit) are represented as a matrix $\mathbf{Y} \in \{0,1\}^{N \times C}$ (where $Y_{ic}=1$ indicates that the $i$-th T-F unit belongs to source $c$). The objective function of Deep Clustering is typically defined as minimizing the Frobenius norm distance between the affinity matrices derived from these two matrices:
\begin{equation}
    \mathcal{L}_{\text{DC}} = ||\mathbf{V}\mathbf{V}^T - \mathbf{Y}\mathbf{Y}^T||_F^2,
\end{equation}
where $\mathbf{V}\mathbf{V}^T$ and $\mathbf{Y}\mathbf{Y}^T$ represent the affinity (inner product) matrices between embedding vectors and between ideal labels, respectively. Since the matrix product $\mathbf{A}\mathbf{A}^T$ is invariant to the column order of matrix $\mathbf{A}$ (i.e., the permutation order of source labels), the loss function $\mathcal{L}_{DC}$ is permutation-invariant. This allows the network to learn discriminative embedding representations $\mathbf{V}$ without knowing the specific correspondence between outputs and targets, thus elegantly solving the label permutation problem. The separation process is completed during the inference phase by applying clustering algorithms (such as K-means, specifying the number of clusters as $C$) to the learned embedding vectors $\mathbf{V}$ to partition T-F units into different clusters, each corresponding to a mask $\hat{\mathbf{M}}_c$ for a separated source. Finally, the estimated signals are obtained by masking the mixed spectrogram $X(t,f)$ and performing inverse STFT: $\hat{\mathbf{s}}_c = \text{iSTFT}(\hat{\mathbf{M}}_c \odot \mathbf{X})$.

Speech separation methods based on deep clustering exhibit diverse forms and applications \cite{isik2016single,luo2017deep,luo2017speaker,chen2017deep,zeghidour2020wavesplit,wang2019deep,wang2018alternative}. The most direct form strictly follows its original definition \cite{isik2016single,chen2017deep}: first, using a deep neural network (typically a recurrent neural network such as BLSTM) to learn embedding vectors for time-frequency units, with the training objective of optimizing the embedding space structure to satisfy the characteristic that elements from the same source cluster together while those from different sources remain separated; then, during the inference phase, applying standard clustering algorithms (such as K-means or spectral clustering) to the embedding vectors output by the network to obtain estimated masks for each source, thereby reconstructing the speech signals. Beyond this basic application, deep clustering has also inspired many improvements and hybrid methods. For instance, the ``Chimera" network \cite{wang2018alternative} and its improved version ``Chimera++" \cite{wang2019deep} combine deep clustering with Mask Inference (MI), setting up two output heads in one network: one head outputs embedding vectors $\mathbf{V}$ for clustering (optimizing $\mathcal{L}_{DC}$), while the other directly predicts separation masks $\hat{M}$ (optimizing mask-related losses $\mathcal{L}_{MI}$, such as mean squared error). The total loss function for this hybrid architecture is typically a weighted sum of both:
\begin{equation}
    \mathcal{L}_{\text{total}} = \alpha \mathcal{L}_{\text{DC}} + (1 - \alpha) \mathcal{L}_{\text{MI}},
\end{equation}
where $\alpha$ is a weighting factor. This approach allows the deep clustering loss $\mathcal{L}_{DC}$ to serve as a regularization means, assisting the mask prediction task, and typically achieves better performance than using either method alone. Another important evolution is the Deep Attractor Network (DAN) \cite{luo2017speaker,chen2017deep}, which borrows the embedding concept from deep clustering but explicitly learns ``attractor" points $\mathbf{a}_c \in \mathbb{R}^D$ representing each source $c$ in the embedding space. These attractors are typically defined as the centroids of embedding vectors for T-F units corresponding to that source. Then, by calculating the distance (e.g., Euclidean distance) between each T-F unit's embedding vector $\mathbf{v}_i$ and various attractors $\mathbf{a}_c$, the unit is assigned to the source represented by the nearest attractor, thereby generating separation masks: $\hat{\mathbf{c}}_i = \arg\min_c ||\mathbf{v}_i - \mathbf{a}_c||^2$. The training objective of DAN is typically to minimize signal reconstruction error, making it an end-to-end method while utilizing the attractor mechanism to maintain permutation invariance. Additionally, researchers have explored different deep clustering objective functions \cite{wang2019deep,zeghidour2020wavesplit} and strategies that combine deep clustering with other techniques (such as signal enhancement post-processing \cite{isik2016single}, iterative phase reconstruction \cite{wang2019deep}, etc.), further expanding the separation framework based on deep clustering ideas.

Deep clustering, as a pioneering speech separation technique, exhibits evident advantages. Its most fundamental contribution lies in providing, for the first time, an effective framework for addressing the label permutation problem in deep learning-based single-channel, speaker-independent speech separation, significantly advancing developments in this field. Additionally, deep clustering demonstrates potential for handling a variable number of sources \cite{hershey2016deep,isik2016single}, as the embedding and clustering framework is not inherently limited to a fixed number of output channels. The number of separated sources can be determined during testing by specifying the number of clusters, and models trained on mixtures with fewer speakers can potentially be applied to separation tasks involving more speakers. However, deep clustering also presents some inherent limitations. Its performance is constrained by the quality of the embedding vectors and the effectiveness of the subsequent clustering algorithm, constituting a two-stage process that may not be end-to-end optimal. The original loss function based on affinity matrices does not directly optimize final speech separation evaluation metrics (such as SDR) \cite{luo2018speaker}, although subsequent research has attempted improvements by introducing end-to-end objectives. Furthermore, the clustering step during inference (especially spectral clustering) may incur high computational costs, and if clustering is performed on short segments, additional steps are needed to resolve cluster label consistency issues across segments.

Despite the subsequent emergence of end-to-end time-domain models (such as TasNet \cite{luo2018tasnet}) and novel architectures incorporating clustering concepts (such as WaveSplit \cite{zeghidour2020wavesplit}) that have achieved superior performance on specific benchmarks, the pioneering idea of deep clustering—utilizing embedding spaces to address permutation invariance—continues to exert profound influence. Future separation models are likely to continue borrowing and developing this embedding learning paradigm, exploring more effective embedding representations, more optimized clustering or assignment mechanisms, and integrating these ideas into more powerful end-to-end trainable frameworks to address more complex and realistic speech separation scenarios, such as those involving reverberation, strong noise, and numerous overlapping speakers.

\subsubsection{Permutation Invariant Training Methods}

\begin{figure}[!t]
\centering
\includegraphics[width=1.0\linewidth]{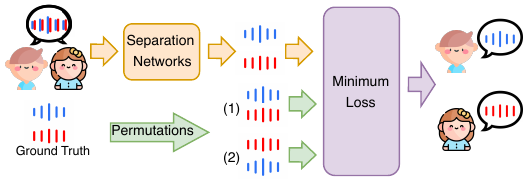}
\caption{The pipeline of the permutation invariant training method.}
\label{fig:pit}
\end{figure}

Unlike deep clustering, Permutation Invariant Training (PIT) resolves this issue through an ingenious strategy: instead of presetting a fixed output order, it considers all $C!$ possible permutations between outputs and targets at each training step \cite{yu2017permutation,kolbaek2017multitalker}, as shown in Fig.~\ref{fig:pit}. Specifically, for a permutation $\sigma \in S_C$ (where $S_C$ is the set of all permutations of $\{1, \dots, C\}$), it calculates the total loss under that permutation. The objective function of PIT is to select the permutation that minimizes the total loss and update the network parameters $\theta$ based on this minimum loss. The loss function is formally expressed as:
\begin{equation}
    \mathcal{L}_\text{PIT} = \min_{\sigma \in S_C} \sum_{i=1}^{C} \mathcal{L}(\mathbf{s}_i, \hat{\mathbf{s}}_{\sigma(i)}),
    \label{eq:pit_loss}
\end{equation}
where $\mathcal{L}(\cdot, \cdot)$ is a loss function measuring the difference between an individual source signal and its estimated signal, such as negative Signal-to-Noise Ratio (SNR) \cite{vincent2006performance} or the commonly used Scale-Invariant Signal-to-Noise Ratio (SI-SNR) \cite{le2019sdr}, and $\sigma(i)$ represents the output index corresponding to the $i$-th target source $\mathbf{s}_i$ under permutation $\sigma$. This strategy directly optimizes separation as the objective, avoiding the label permutation problem and significantly improving model performance on speaker-independent separation tasks \cite{yu2017permutation}.

The original PIT \cite{yu2017permutation} computes optimal permutations at the frame or segment level, which can lead to inconsistent speaker-to-channel assignments across a long utterance, disrupting temporal continuity. To mitigate this, several extensions have been proposed, evolving PIT to handle longer contexts, variable speaker counts, and complex scenarios like meetings. Utterance-level PIT (uPIT) \cite{kolbaek2017multitalker} extends permutation selection to the entire utterance, identifying a single optimal mapping that ensures consistent speaker assignments throughout the signal. This promotes learning of long-term dependencies and speaker traits. Key advantages include improved continuity and performance in fixed-speaker scenarios; however, it assumes the number of speakers $K$ does not exceed output channels $C$ ($K \le C$), limiting flexibility in dynamic environments. Building on this, Graph-PIT \cite{von2021graph} relaxes the $K \le C$ constraint, allowing more speakers overall as long as simultaneously active ones do not exceed $C$. It models speech segments as a graph, with nodes for segments and edges for temporal overlaps, then frames permutation as a graph coloring problem to minimize loss while ensuring non-overlapping segments can share channels. This is particularly useful for conference-like recordings with intermittent speakers. Advantages include scalability to variable and larger speaker sets; drawbacks involve higher computational complexity due to graph optimization, potentially scaling poorly with many nodes. Alternatively, One-and-Rest PIT (OR-PIT) \cite{takahashi2019recursive} employs a recursive approach, iteratively separating one speaker and feeding the residual mixture back into the network until all are extracted. The recursion depth adapts to the (unknown) number of speakers. This offers flexibility for varying $K$ without fixed $C$ assumptions. Strengths lie in handling unknown speaker counts and recursive refinement; limitations include potential error accumulation across iterations and increased inference time. These extensions highlight PIT's evolution from local to global and adaptive handling, addressing real-world challenges like speaker variability, but often at the cost of added complexity or assumptions.

PIT successfully addresses the fundamental challenge of label permutation, enabling end-to-end deep learning models to effectively tackle speaker-independent speech separation tasks and achieve breakthrough performance improvements \cite{wang2023tf,zhao2023mossformer,zhao2024mossformer2}. Models trained with PIT and its variants (such as uPIT) demonstrate good generalization capabilities, able to separate voices of speakers not present in the training set, and even generalize across languages, indicating that the models learn relatively universal acoustic separation cues. As a training strategy, PIT can be flexibly combined with various complex neural network architectures (such as RNN \cite{yu2017permutation}, LSTM \cite{kolbaek2017multitalker,luo2018tasnet,luo2020dual,chen2020dual}, CNN \cite{luo2019conv,tzinis2020sudo,hu2021speech,li2022efficient,xu2024tiger}, and Transformers \cite{wang2023tf,subakan2021attention,zhao2023mossformer,zhao2024mossformer2}). For uPIT, the additional permutation computation overhead during training is typically acceptable relative to complex deep models, while there is no extra computational burden during the inference phase. However, PIT methods also have some disadvantages. Basic frame-level PIT may lead to inconsistent correspondence between output channels and speakers over time; although uPIT alleviates this issue, enforcing globally consistent permutations for very long audio might be overly restrictive. The main limitation lies in computational complexity: PIT and uPIT need to evaluate $C!$ permutations, which becomes impractical when the number of speakers $C$ increases (e.g., greater than 3 or 4) as the computational cost grows dramatically \cite{zhu2021multi,nachmani2020voice}. While Graph-PIT handles more potential speakers through graph modeling, it introduces the complexity of graph construction and coloring algorithms. OR-PIT, though avoiding the $C!$ computation, may employ a suboptimal recursive/iterative separation approach and potentially suffer from error accumulation problems.

Future research should focus on addressing key limitations of PIT-based methods. This includes developing efficient permutation handling for long audio and dynamic speaker scenarios \cite{chen2020continuous} (e.g., via online processing or segmented PIT), and reducing computational complexity for more than three speakers through approximation algorithms or hybrid approaches like clustering guidance \cite{zeghidour2020wavesplit}. Additionally, improving robustness in noisy, reverberant environments and enabling end-to-end optimization with downstream tasks such as ASR \cite{narayanan2014investigation} and speaker diarization \cite{raj2021integration} will enhance practical applicability.

\section{Architectures}
\label{sec:system}
\begin{figure}[!t]
\centering
\includegraphics[width=1.0\linewidth]{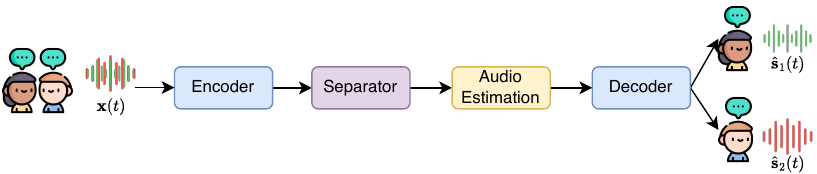}
\caption{Overall pipeline of speech separation.}
\label{fig:overall-ss}
\end{figure}

In the field of speech separation, the encoder-separator-audio estimation-decoder architecture is the unique framework, as illustrated in Fig.~\ref{fig:overall-ss}. The encoder is responsible for converting the raw audio signal into a more informative representation, the separator processes these representations to isolate different sound sources, and the decoder converts the separated representations back into time-domain signals. The advantage of this architecture lies in its ability to decompose the complex speech separation task into three relatively independent modules—feature extraction, feature separation, and signal reconstruction—which facilitates model design and optimization. This section presents a comprehensive chronological analysis of speech separation models as categorized in Table \ref{tab:sorted_models}, offering insights into the evolution of various encoder-separator-audio estimation-decoder architectures across different methodological paradigms. In this section, unless otherwise specified, we restrict our discussion to scenarios where the number of speakers is known.

\begin{table*}[!t]  
\centering
\renewcommand{\arraystretch}{1.0}  
\caption{Overview of speech separation models categorized by encoder/decoder architecture (STFT/Convolution/Pre-trained), separator (RNN/CNN/Attention/Mixture-based), and audio estimated method (mask-based vs. mapping-based).}
\label{tab:sorted_models}
\begin{tabular}{l|p{0.25\textwidth} p{0.25\textwidth} p{0.20\textwidth}} 
\toprule
\textbf{Methods} & \textbf{Encoder/Decoder} & \textbf{Separator} & \textbf{Audio Estimation} \\
\midrule
DPCL\cite{hershey2016deep}                  & STFT         & RNN-Based       & mask    \\
PIT\cite{yu2017permutation}                 & STFT         & RNN-Based       & mask    \\
uPIT-BLSTM\cite{kolbaek2017multitalker}     & STFT         & RNN-Based       & mask    \\
DANet\cite{chen2017deep}                    & STFT         & RNN-Based       & mask    \\
ADAN\cite{luo2017speaker}                   & STFT         & RNN-Based       & mask    \\
Chimera++\cite{wang2018alternative}         & STFT         & RNN-Based       & mask    \\
TaSNet\cite{luo2018tasnet}                  & Convolution  & RNN-Based       & mask    \\
SSGAN-PIT\cite{chen2018permutation}         & STFT         & RNN-Based       & mask    \\
CBLDNN-GAT\cite{li2018cbldnn}               & STFT         & RNN-Based       & mask    \\
Wave-UNet\cite{stoller2018wave}             & Convolution  & CNN-Based       & mapping \\
Chimera++ sign\cite{wang2019deep}           & STFT         & Mixture-Based     & mask    \\
Conv-TasNet\cite{luo2019conv}                & Convolution  & CNN-Based       & mask    \\
Deep CASA\cite{liu2019divide}               & STFT         & CNN-Based       & mask    \\
OR-PIT\cite{takahashi2019recursive}         & Convolution  & RNN-Based       & mask    \\
WaveSplit\cite{zeghidour2020wavesplit}      & Convolution  & RNN-Based       & mapping \\
DPTNet\cite{chen2020dual}                   & Convolution  & Mixture-Based     & mask    \\
Conv-TasSAN\cite{deng2020conv}              & Convolution  & CNN-Based       & mask    \\
DPRNN\cite{luo2020dual}                     & Convolution  & RNN-Based       & mask    \\
VSUNOS\cite{nachmani2020voice}              & Convolution  & RNN-Based       & mapping \\
Two-Step TCN\cite{tzinis2020two}            & Convolution  & CNN-Based       & mask    \\
SudoRM-RF\cite{tzinis2020sudo}              & Convolution  & CNN-Based       & mask    \\
MixIT\cite{wisdom2020unsupervised}          & Convolution  & CNN-Based       & mapping \\
FurcaNeXt\cite{zhang2020furcanext}          & Convolution  & CNN-Based       & mapping \\
TS-MixIT\cite{zhang2021teacher}             & Convolution  & CNN-Based       & mapping \\
SepFormer\cite{subakan2021attention}        & Convolution  & Attention-Based & mask    \\
MSGT-TasNet\cite{zhao2021multi}             & Convolution  & Attention-Based & mask    \\
Multi-Decoder Dprnn\cite{zhu2021multi}      & Convolution  & RNN-Based       & mapping \\
DPTCN-ATPP\cite{zhu2021dptcn}               & Convolution  & CNN-Based       & mask    \\
Unknow-SS\cite{chazan2021single}            & Convolution  & CNN-Based       & mapping \\
A-FRCNN\cite{hu2021speech}                  & Convolution  & Mixture-Based     & mask    \\
Sandglasset\cite{lam2021sandglasset}        & Convolution  & Mixture-Based     & mask    \\
CDGAN\cite{li2021generative}                & STFT         & RNN-Based       & mapping \\
SSL-SS\cite{huang2022investigating}         & STFT         & RNN-Based       & mask    \\
SkiM\cite{li2022skim}                       & Convolution  & RNN-Based       & mapping \\
TDANet\cite{li2022efficient}                & Convolution  & Mixture-Based & mask    \\
MTDS\cite{qian2022efficient}                & Convolution  & RNN-Based       & mask    \\
QDPN\cite{rixen2022qdpn}                    & Convolution  & Mixture-Based     & mapping \\
SFSRNet\cite{rixen2022sfsrnet}              & Convolution  & Attention-Based & mask    \\
TFPSNet\cite{yang2022tfpsnet}               & STFT         & Attention-Based & mask    \\
SepEDA\cite{chetupalli2022speech}           & Convolution  & Mixture-Based     & mask    \\
SepDiff\cite{chen2023sepdiff}               & STFT         & CNN-Based       & mapping \\
S4M\cite{chen2023neural}                    & Convolution  & Mixture-Based     & mask    \\
HuBERT\cite{fazel2023cocktail}              & Pre-trained Model    & Attention-Based & mapping \\
pSkiM\cite{li2023predictive}                & Convolution  & RNN-Based       & mapping \\
PGSS\cite{li2023pgss}                       & STFT         & Mixture-Based     & mapping \\
Separate And Diffuse\cite{lutati2023separate} & STFT       & Attention-Based & mapping \\
DiffSep\cite{scheibler2023diffusion}        & STFT         & CNN-Based       & mapping \\
TF-GridNet\cite{wang2023tf}                 & STFT         & Mixture-Based     & mapping \\
UNSSOR\cite{wang2023unssor}                 & STFT         & Attention-Based & mapping \\
MossFormer\cite{zhao2023mossformer}         & Convolution  & Attention-Based     & mask    \\
ReSepFormer\cite{della2024resource}         & Convolution  & Attention-Based & mask    \\
Conv-TasNet GAN\cite{lakandri2024exploring} & Convolution  & CNN-Based       & mask    \\
SepTDA\cite{lee2024boosting}                & Convolution  & Mixture-Based     & mapping \\
SPMamba\cite{li2024spmamba}                 & STFT         & Mixture-Based & mapping \\
Fast-GeCo\cite{wang2024noise}               & STFT         & CNN-Based       & mapping \\
DIP\cite{wang2024speech}                    & Pre-trained Model    & Mixture-Based     & mask    \\
TIGER\cite{xu2024tiger}                     & STFT         & Mixture-Based     & mask    \\
CodecSS\cite{yip2024speech}                & Pre-trained Model    & Attention-Based & mapping \\
TCodecSS\cite{yip2024towards}               & Pre-trained Model    & Attention-Based & mapping \\
MossFormer2\cite{zhao2024mossformer2}       & Convolution  & Attention-Based     & mask    \\
EDSep\cite{dong2025edsep}                   & STFT         & CNN-Based       & mapping \\
\bottomrule
\end{tabular}
\end{table*}

\subsection{Encoder \& Decoder}

\begin{figure*}[!t]
\centering
\includegraphics[width=1.0\linewidth]{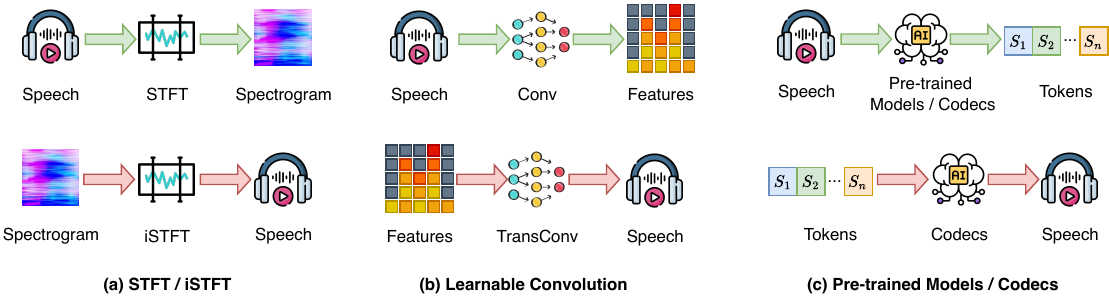}
\caption{The process of feature extraction by the encoder and waveform reconstruction by the decoder in speech separation.}
\label{fig:encoder-decoder}
\end{figure*}

Currently, the mainstream encoders and decoders can be broadly categorized into the following types, as shown in Fig.~\ref{fig:encoder-decoder}. The first category relies on the short-time Fourier transform (STFT) and its inverse (iSTFT). In this approach, the STFT converts a time-domain signal into a time-frequency domain representation, leveraging a two-dimensional depiction of time and frequency to differentiate the characteristics of various sound sources. This method has been widely used in early studies as well as in some current research \cite{hershey2016deep,chen2017deep,luo2017speaker,wang2023tf,xu2024tiger,li2024spmamba}. Its advantages lie in its clear physical interpretation, ease of understanding and implementation, and compatibility with many traditional signal processing techniques. However, as STFT/iSTFT are fixed, non-learnable transforms, their time-frequency resolution is inherently limited by the uncertainty principle related to window function and stride selection. Consequently, they may not provide the optimal feature representation for separation tasks, and a larger window length—often required in low-latency scenarios—can introduce significant delay \cite{li2022skim}. Moreover, when reconstructing the time-domain signal using iSTFT, typically only the magnitude spectrum is employed \cite{hershey2016deep,chen2017deep,luo2017speaker}, while the phase information is often estimated via the Griffin-Lim algorithm \cite{griffin1984signal} or directly taken from the mixed signal, potentially leading to degradation in the quality of the reconstructed signal. Recently, certain methods have converted the complex-domain features obtained from STFT into the real domain so that the separation network can jointly estimate the real and imaginary components of the target signal, thereby enabling a joint estimation of magnitude and phase information \cite{xu2024tiger,wang2023tf,luo2023music}.

The second category is based on learnable convolutional neural networks (CNNs). For example, one-dimensional convolutions (1D Conv) can be applied directly in the time domain to encode the waveform, using multiple convolutional layers and downsampling operations to extract features. Corresponding decoders then employ one-dimensional transposed convolutions (1D Transposed Conv) or upsampling operations to reconstruct the waveform \cite{luo2018tasnet,luo2019conv}. This method integrates the processes of feature extraction and reconstruction into an end-to-end learning framework, allowing the network to automatically learn an optimal basis of feature representations according to the requirements of the separation task \cite{luo2020dual,chen2020dual,subakan2021attention,tzinis2020sudo,hu2021speech,li2022efficient}. Its advantages include high flexibility and the ability to directly optimize time-domain separation metrics (such as SI-SNR \cite{le2019sdr}), thereby circumventing the phase estimation issues and resolution limitations of the STFT. The learnable encoder and decoder used in Conv-TasNet \cite{luo2019conv} are typical examples within this direction. Nevertheless, the learned features often lack an explicit physical interpretation, leading to relatively low model interpretability; moreover, a larger corpus of training data may be required to learn effective feature transformations \cite{michelsanti2021overview}. Future research might achieve breakthroughs either by designing more efficient convolutional structures to capture long-term dependencies or by incorporating physical priors into the feature learning process.

The third category involves the use of pre-trained models or neural network-based codecs as encoders. In recent years, with the advancement of self-supervised learning, models such as Wav2Vec 2.0 \cite{baevski2020wav2vec} and HuBERT—pre-trained \cite{hsu2021hubert} on large-scale unlabeled speech data—have demonstrated the ability to extract rich acoustic and speech information. By employing the encoder component of these pre-trained models or their extracted embeddings as the encoder in a speech separation system \cite{neri2021unsupervised,wang2024speech}, one can leverage their powerful representational capabilities to potentially enhance generalization performance in complex acoustic scenarios. Similarly, neural network-based codecs (e.g., EnCodec \cite{defossez2022high}, SoundStream \cite{zeghidour2021soundstream}) can compress speech into low-bitrate discrete or continuous representations, while still enabling high-quality reconstruction. Utilizing such codec-based encoders for feature extraction, along with appropriate decoders for reconstruction, has emerged as a new research direction \cite{yip2024speech,yip2024towards}. The advantages of this approach include leveraging the robust feature extraction capability afforded by large-scale pre-training, which may provide enhanced robustness to noise and reverberation. On the other hand, pre-trained models often have a large number of parameters and high computational complexity, rendering them unsuitable for devices with limited resources \cite{yoon2022hubert}. Moreover, discrepancies between the objectives of pre-training tasks (such as mask prediction or contrastive learning) and those of the speech separation task may necessitate careful interface design or fine-tuning \cite{wang2024noise}. Additionally, the use of neural network-based codecs may introduce compression artifacts. Future work is expected to focus on efficiently transferring pre-trained knowledge to the domain of speech separation, designing lightweight pre-trained models or codecs, and developing pre-training strategies specifically tailored for separation tasks.

\subsection{Separator Architectures}

\begin{figure*}[!t]
\centering
\includegraphics[width=1.0\linewidth]{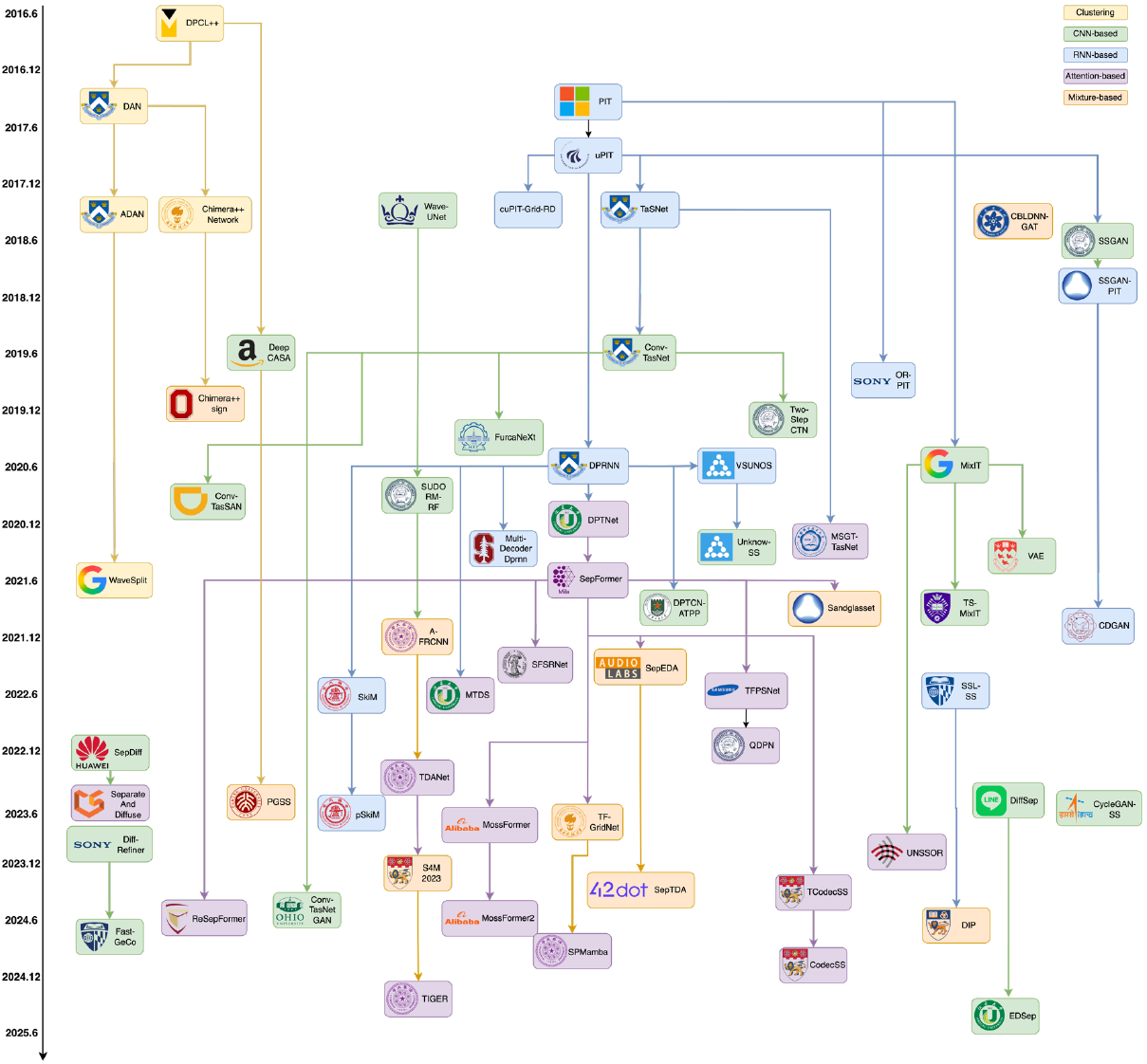}
\caption{Overview of the evolution of separator architectures in speech separation. The figure illustrates the development from single-structure models to hybrid architectures, highlighting the emergence of representative paradigms including RNN-based, CNN-based, Attention-based, and Mixture-based approaches. Note: As many publications involve authors from multiple institutions, the affiliation for each method is determined by the corresponding author's institution, or the first author's institution if a corresponding author is not specified.}
\label{fig:all_model}
\end{figure*}

The separator constitutes the core component of speech separation models, with its architectural design directly determining the model’s capacity to extract and model target speech characteristics from mixed signals, thereby affecting overall separation performance. Throughout the evolution of speech separation techniques (as shown in Figure \ref{fig:all_model}), separator architectures have progressed from single-structure designs to hybrid integrations, giving rise to several mainstream paradigms. These can be categorized into RNN-based, CNN-based, Attention-based, and mixed architecture (Mixture-based) approaches. In this section, we provide a detailed description of these four principal separator architectures, discussing their core concepts, representative models, current challenges, and future directions.

\subsubsection{RNN-Based Methods}

Recurrent Neural Networks (RNNs) have occupied an important position in the field of speech separation due to their inherent sequence modeling capabilities. Speech signals are essentially time-series data, where temporal dependencies are crucial for understanding and separating mixed sound sources. RNNs, especially Long Short-Term Memory networks (LSTMs) and Gated Recurrent Units (GRUs), can effectively capture and model these long-distance temporal dependencies, making them ideal choices for handling speech separation tasks \cite{hershey2016deep, yu2017permutation}. With the development of deep learning techniques, the application of RNNs in speech separation has gradually expanded from initial frequency-domain feature processing to time-domain direct modeling, further improving separation performance and processing efficiency \cite{luo2018tasnet, li2022skim,wang2023tf,li2024spmamba,chen2020dual}.

\begin{figure}[!t]
\centering
\includegraphics[width=1.0\linewidth]{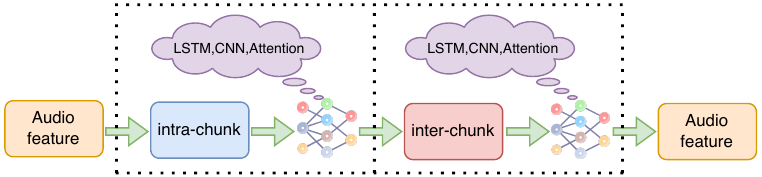}
\caption{Overall pipeline of DPRNN.}
\label{fig:overall-dprnn}
\end{figure}

RNN-based speech separation methods still face several inherent limitations that hinder their performance and applicability in real-world scenarios. One major issue is the vanishing or exploding gradient problem during training, particularly in standard RNNs, which makes it difficult to capture very long-range dependencies in speech signals \cite{subakan2021attention}. This is exacerbated in speech separation tasks involving extended audio sequences, where temporal context spans minutes or more, leading to degraded separation quality for distant sound sources or overlapping speakers \cite{quan2024spatialnet,quan2024multichannel}. Another significant drawback is the high computational burden, as RNNs process sequences sequentially, resulting in linear time complexity with respect to sequence length \cite{gu2023mamba,lei2017simple}. This makes them inefficient for long inputs, often requiring substantial memory and processing power, which is problematic for resource-constrained devices like mobile phones or edge computing systems. Additionally, RNNs struggle with real-time requirements in practical applications, such as live conversations or streaming audio, due to their non-parallelizable nature and high latency in causal (online) processing modes. These limitations not only reduce separation accuracy in complex, noisy, or multi-speaker environments but also limit scalability, making RNNs less competitive against more parallel architectures like Transformers in certain scenarios \cite{subakan2021attention, luo2020dual}. Addressing these challenges is crucial for advancing RNNs in speech separation, as they represent key research opportunities for improving efficiency, robustness, and adaptability.

To mitigate these drawbacks, researchers have introduced innovative strategies that enhance RNNs without sacrificing their core temporal modeling strengths. For instance, DPRNN divides long sequences into shorter chunks and employs a dual-path structure for intra-chunk and inter-chunk processing, effectively alleviating the long-range dependency and computational issues \cite{luo2020dual}. Similarly, SkiM incorporates a skip-memory mechanism to reduce redundant computations by approximately 75\%, maintaining separation performance while lowering overhead \cite{li2022skim}. Building on this, Predictive SkiM integrates contrastive predictive coding with local context encoding-decoding to boost accuracy and real-time capabilities in causal modeling \cite{li2023predictive}. These approaches improve RNNs' handling of complex scenarios like multi-speaker dialogues, providing a foundation for more efficient and robust systems. Building on these improvements, there are still many potential directions for the development of RNNs in speech separation. For example, combining RNNs with other architectures, such as convolutional neural networks \cite{luo2019fasnet}, attention mechanisms \cite{wang2023tf}, or diffusion models, could create hybrid models that leverage complementary strengths for better local detail capture and global context understanding. Additionally, optimizing techniques like contrastive predictive coding, skip-memory mechanisms, and autoregressive strategies may further reduce latency and computational costs, enabling deployment on mobile or low-power devices. For ultra-long sequences or varying speaker counts, integrating RNNs with advanced frameworks like Transformers \cite{vaswani2017attention} or State-Space Models \cite{gu2023mamba,li2024spmamba} holds promise for groundbreaking improvements.

\subsubsection{CNN-Based Methods}

Convolutional Neural Networks (CNNs) have become one of the primary network architectures used in speech separation models due to their inherent feature extraction capabilities and efficiency in capturing local temporal dependencies in audio signals. The widespread adoption of this technology primarily stems from the rise of end-to-end models that operate directly in the time domain \cite{luo2019conv}, circumventing the complexity of phase reconstruction in traditional time-frequency methods \cite{tzinis2020sudo,hu2021speech,deng2020conv,pariente2020filterbank}. CNNs provide a powerful mechanism for learning complex representations from raw waveforms, enabling models to directly map mixed audio signals to separated sources. Furthermore, advancements in CNN architectures such as depthwise separable convolutions and dilated convolutions have made it possible to construct efficient models that simultaneously capture both short-term and long-term temporal dynamics, perfectly suited for the complex task of audio source separation \cite{luo2019conv}.

CNN-based speech separation techniques present a diverse landscape, encompassing various architectural designs that effectively utilize convolutional layers. Among these, encoder-decoder structures are particularly prominent, such as Wave-U-Net \cite{stoller2018wave} and SUDO RM-RF \cite{tzinis2020sudo}, inspired by U-Net \cite{ronneberger2015u} from the image segmentation domain. These models progressively downsample the input through an encoder to extract hierarchical features, then reconstruct separated signals via a decoder with upsampling, typically incorporating skip connections to preserve fine-grained details. Another important category includes models like Conv-TasNet \cite{luo2019conv} and DPTCN-ATPP \cite{zhu2021dptcn}, which utilize stacked one-dimensional convolutional blocks to learn feature representations and generate separation masks. These models frequently employ dilated convolutions to expand the temporal receptive field and capture multi-scale features through parallel processing paths \cite{xu2024tiger}, thereby enhancing their ability to model complex time-frequency patterns in speech. Additionally, attempts like learnable fbank to explore learnable filterbanks within the CNN framework demonstrate efforts to optimize front-end processing of audio signals \cite{pariente2020filterbank}, moving beyond fixed filterbanks toward data-driven representations that can improve separation performance, particularly in noisy environments.

Despite the significant achievements of CNN methods in speech separation tasks, they still face several critical challenges that limit their performance and generalization. Foremost, CNNs often struggle to adequately capture long-range dependencies in audio signals while maintaining sensitivity to local features, as their convolutional kernels are inherently limited in receptive field size, potentially leading to suboptimal modeling of global temporal contexts in complex mixtures \cite{hu2021speech}. Secondly, in encoder-decoder structures like Wave-U-Net and SUDO RM-RF, the upsampling process can introduce artifacts or distortions, such as aliasing or phase inconsistencies, which degrade reconstruction quality and affect the perceptual fidelity of separated speech \cite{pariente2020filterbank}. Furthermore, direct time-domain modeling imposes high demands on network parameter convergence and feature matching, often resulting in training instability, sensitivity to hyperparameters, and poor robustness in noisy or reverberant environments \cite{tzinis2020sudo}. These limitations highlight key research gaps: for instance, how to extend receptive fields without exponentially increasing computational costs, how to mitigate reconstruction errors in upsampling, and how to stabilize training for real-world deployment. Addressing these issues is crucial for advancing CNN-based models, as they represent fertile ground for future innovations in speech separation. To mitigate these challenges, researchers have explored several strategies. For example, models like Conv-TasNet incorporate non-linear encoders and parallel multi-scale separation modules to better integrate local and global information, while improved training strategies such as multi-task learning enhance model robustness \cite{tzinis2020sudo,hu2021speech,li2022efficient}. Similarly, DPTCN-ATPP employs dilated convolutions and dual-path parallel encoding to expand receptive fields and reduce instability in time-domain processing \cite{zhu2021dptcn}.

The development of CNNs in speech separation will primarily focus on further optimization of network structures and training mechanisms to overcome the aforementioned limitations. On one hand, the fusion of multi-scale, multi-resolution information, attention mechanisms, and adaptive filterbanks (as proposed in Learnable fbank) is expected to improve separation effects and noise resistance performance \cite{pariente2020filterbank}. On the other hand, leveraging emerging technologies such as Transformers, dual-path modeling, and self-supervised learning to combine CNNs with other deep models may enable more precise capture of long-distance dependencies and handling of complex noise scenarios; for instance, Two-Step TCN provides insights for low-latency and real-time applications through a strategy of first learning latent spaces before separation \cite{tzinis2020two}. Finally, the design of lightweight CNN models will receive increasing attention to meet requirements for real-time processing and deployment on resource-constrained devices \cite{hu2021speech,li2022efficient,xu2024tiger}.

\subsubsection{Attention-Based Methods}
In recent years, with the success of Transformer in sequence modeling tasks \cite{vaswani2017attention,achiam2023gpt,liu2024deepseek}, attention-based speech separation methods have gradually become a research hotspot \cite{wang2023tf,zhao2023mossformer,zhao2024mossformer2,chen2020dual,subakan2021attention}. Although these methods differ in specific implementations, they generally share two significant characteristics: First, they utilize self-attention mechanisms to enable direct interaction of global information, breaking the limitations of traditional RNNs or CNNs in capturing long-distance dependencies \cite{subakan2021attention,zhao2023mossformer,zhao2024mossformer2}; Second, through segmentation or dual-path \cite{wang2023tf}, multi-scale strategies \cite{zhao2021multi}, they consider both local details and effectively integrate long temporal information, thereby better accomplishing the separation of mixed speech signals. The Transformer architecture also allows parallel processing, avoiding the inherent sequential computation limitations in RNNs, which significantly improves the training and inference efficiency of models\cite{zhao2021multi}. Attention-based models have achieved state-of-the-art separation performance on standard datasets such as WSJ0-2mix, demonstrating their effectiveness in the field of speech separation \cite{zhao2023mossformer,zhao2024mossformer2}.

It is crucial to distinguish that the ``attention" mechanism central to these computational models is conceptually different from the top-down cognitive attention discussed in Section \ref{sec:cocktail-party}. The latter refers to a selective, goal-driven process in the human brain, where a listener actively focuses on a target speaker using prior knowledge \cite{awh2012top} (e.g., the speaker's voice characteristics). In contrast, the self-attention mechanism in Transformers can be understood as a form of bottom-up attention. It is a data-driven computational process that dynamically calculates the relevance of every element in an input sequence to every other element. This ``focus" is not guided by an external, pre-defined goal but rather emerges from the statistical patterns learned from the data itself. Thus, it is a mechanistic weighting scheme for feature integration, distinct from the brain's high-level selective auditory processing.

Attention-based speech separation methods can be categorized into several types based on their architecture and application of attention mechanisms. One approach is the dual-path processing strategy, such as the Dual-Path Transformer Network (DPTNet) \cite{luo2020dual} and SepFormer \cite{subakan2021attention}, which segment the input signal along the time axis and apply Transformers to capture local temporal features (Intra-Transformer) and long-term dependencies across blocks (Inter-Transformer). This dual-path structure enables the model to effectively process long sequences while maintaining sensitivity to short-term patterns. Another approach introduces special attention mechanisms to reduce computational complexity, such as MSGT-TasNet \cite{zhao2021multi}, which employs group self-attention mechanisms and multi-scale fusion to reduce complexity while maintaining modeling capability for long-term dependencies. Sandglasset \cite{lam2021sandglasset} introduces an hourglass-shaped architecture with multi-granularity features, where the temporal granularity of features gradually coarsens in the first half of the network and then gradually refines toward the original signal level. This unique structure allows the model to capture both broad and detailed temporal information. TFPSNet \cite{yang2022tfpsnet} combines time-frequency domain analysis, where multi-path blocks are designed not only to scan individually on time and frequency domains but also include blocks that scan along the time-frequency domain diagonal. 
The MossFormer series models \cite{zhao2023mossformer,zhao2024mossformer2} adopt a gated single-head Transformer architecture and a convolution-enhanced joint self-attention mechanism, through joint local and global self-attention mechanisms, performing full-computation self-attention on local blocks and linearized low-cost self-attention across the entire sequence.

However, these methods are not without significant drawbacks, which pose substantial challenges for practical applications and represent key areas for future research. The most prominent issue is the high computational complexity, especially when processing long sequences. In the standard Transformer's self-attention mechanism, the complexity grows quadratically with sequence length (i.e., $O(n^2)$ where n is the sequence length), making it computationally prohibitive for speech sequences that can extend to thousands of frames \cite{subakan2021attention}. This not only increases training and inference times but also demands substantial memory resources, limiting deployment on resource-constrained devices like edge hardware \cite{wang2023tf,subakan2021attention,chen2020dual,yang2022tfpsnet}. Additionally, attention mechanisms can suffer from overparameterization, leading to overfitting in low-data scenarios, and may struggle with capturing fine-grained local features without additional hybrid components \cite{li2022efficient,xu2024tiger}. These limitations highlight critical research gaps: how to scale attention for ultra-long sequences without exponential resource demands, how to mitigate energy inefficiency in real-time systems, and how to balance global modeling with local precision in dynamic noise environments.

To mitigate these challenges, researchers have explored various improvement strategies. For instance, multi-group self-attention (MSGT) divides the input into smaller groups to compute intra-group correlations, maintaining complexity within acceptable bounds as long as group sizes are controlled \cite{zhao2021multi}. Sparse Transformers employ sparse attention patterns to reduce unnecessary computations, while hierarchical attention networks and segment-level recursive mechanisms enable learning of dependencies beyond fixed lengths \cite{dai2019transformer}.

\subsubsection{Mixture-Based Methods}

Speech separation methods based on mixed architectures typically integrate multiple network structures to address different subtasks such as multi-scale information extraction \cite{xu2024tiger,li2022efficient,hu2021speech} and time-frequency features \cite{wang2023tf,li2018cbldnn,wang2019deep}. The common characteristic is the incorporation of various modules including CNN, RNN, or Transformer to fully utilize local spectral features and global temporal context, thereby achieving complementary advantages in performance and convergence stability \cite{xu2024tiger,wang2023tf}. These methods adopt a ``hybrid" strategy because single networks often struggle to simultaneously account for spectral details and long-term dependencies, whereas mixed architectures can balance feature extraction and temporal modeling to meet comprehensive requirements for separation quality, real-time processing, and computational resources across different application scenarios.

Mixed architecture approaches encompass various types, with design philosophies primarily reflected in the ``mixing" and information interaction between different modules or branches. For instance, CBLDNN-GAT \cite{li2018cbldnn} achieves enhanced separation effects through hierarchical fusion of convolutional layers and bidirectional LSTM, combined with generative adversarial training to fully utilize log-mel filterbank and pitch information. The Sign Prediction Network \cite{wang2019deep} integrates clustering-based, RNN, and CNN techniques while employing geometric constraints to specifically address phase sign ambiguity issues arising from accurate amplitude estimation. Additionally, A-FRCNN \cite{hu2021speech} implements an asynchronous update strategy with bottom-up, top-down, and lateral connections to achieve collaborative processing of multi-temporal scale information, effectively compressing model parameters while improving separation performance. TF-GridNet \cite{wang2023tf} features a multi-path structure comprising intra-frame spectral modules, sub-band temporal modules, and full-frequency self-attention modules, not only achieving complex spectral mapping but also ensuring that separated speech strictly satisfies energy conservation requirements of the mixed signal through novel loss terms.

Mixed architecture methods face significant challenges that limit their practical deployment. The primary issue is the substantial computational overhead introduced by combining multiple architectural components. For instance, TF-GridNet alternates between bidirectional LSTM and full-frequency self-attention mechanisms, resulting in extremely high computational resource requirements that hinder real-time deployment or implementation on edge devices \cite{li2022efficient}. The model parameter explosion is another critical challenge—these hybrid approaches typically combine convolutional, recurrent, and self-attention modules to capture both local spectral and global temporal information, leading to models with hundreds of millions of parameters. Additionally, training complexity poses significant difficulties. Hybrid models based on generative adversarial training suffer from unstable convergence and mode collapse issues inherent to GANs, making the training process complex and difficult to debug \cite{bang2021mggan}. The multi-task learning frameworks commonly employed in these architectures further complicate the optimization landscape, requiring careful hyperparameter tuning and often leading to suboptimal trade-offs between different objectives. These computational and training challenges severely restrict the applicability of mixed architecture methods in resource-constrained scenarios and real-world deployments.

Recent research has explored lightweight alternatives to address these limitations. State space models (SSM) have emerged as one potential direction, replacing traditional RNN or Transformer architectures by modeling input signals as linear ordinary differential equations \cite{li2024spmamba,chen2023neural}. Some approaches also investigate the integration of auditory mechanisms, such as using pitch contours as prior information for conditional generative networks \cite{wang2023unssor,zhang2019image}. However, these solutions remain largely experimental and have yet to demonstrate consistent performance improvements across diverse acoustic conditions.

\subsection{Audio Estimation Methods}

\subsubsection{Masking Methods}
In the field of speech separation, mask-based approaches \cite{hershey2016deep,luo2018tasnet} have been pivotal in the development of the encoder-separator-decoder architecture. In this setup, the model's goal is to produce one or more masks that, when applied to the high-dimensional representation output by the encoder, separates individual source signals from the original mixed signal. Each mask corresponds to a specific source, and when element-wise multiplied with the encoded representation of the mixed signal, it extracts the desired source component or suppresses the noise. This method relies on the ability of deep learning models to learn to extract useful features from the mixed signal and use these features to generate effective masks. Under this framework, the overall formula of the model is as follows:
\begin{equation}
\begin{aligned}
    \mathbf{W} &= \text{Encoder}(\mathbf{x}), \\
    \mathbf{M} &= \text{Separator}(\mathbf{W}), \\
    \hat{\mathbf{s}}_i &= \text{Decoder}(\mathbf{M}_i \odot \mathbf{W}), \quad \text{for each source } i,
\end{aligned}
\end{equation}
where $\mathbf{x}$ is the input of original speech signal, $\mathbf{W}$ is the high-dimensional representation output by the encoder, $\mathbf{M}$ is the mask which is to extract specific sources or to suppress noise, and $\odot$ denotes element-wise multiplication (also known as the Hadamard product or pointwise multiplication).

Mask-based methods demonstrate significant advantages in practical applications, offering intuitive interpretability as they can be viewed as selective filters that transparently show how the model differentiates and extracts target source signals \cite{hershey2016deep}. However, masking operations may lead to irreversible information loss, particularly when mask values are too low or zero, potentially causing important signal components to be erroneously suppressed \cite{wang2023tf}. Another common issue is that signals processed with masks might introduce artificial artifacts during the decoding process, affecting the naturalness and quality of the final output. Considering these limitations of masking methods, researchers in recent years have focused on developing more advanced separator structures and optimization strategies to extract more discriminative signal representations, thereby maintaining the advantages of masking methods while mitigating their drawbacks \cite{zhao2023mossformer,zhao2024mossformer2}.

\subsubsection{Mapping Methods}
In mapping-based methods \cite{wang2023tf,dong2025edsep}, the goal is not to generate masks but to directly map (or transform) from the encoded representation of the mixed signal to the representations of the individual sources. This approach typically involves a direct mapping function that learns how to recover the signals of independent sources from the mixed signal over time. A major advantage of mapping-based methods is that they can directly optimize the feature space used for separation, rather than indirectly operating through masks.The overall formula for a mapping-based method can be stated as:
\begin{equation}
\begin{aligned}
\mathbf{W} &= \text{Encoder}(\mathbf{x}), \\
\mathbf{S}_i &= \text{Separator}(\mathbf{W}), \quad \text{for each source } i, \\
\hat{\mathbf{s}}_i &= \text{Decoder}(\mathbf{S}_i), \quad \text{for each source } i.
\end{aligned}
\end{equation}

A core advantage of mapping methods lies in their ability to directly optimize the feature space used for separation, avoiding the indirect operations of mask-based methods \cite{wang2023tf,dong2025edsep}. This direct transformation mechanism can theoretically learn more complex source signal disentanglement approaches and potentially circumvent signal distortion or artifact issues caused by imprecise masks, as they do not rely on element-wise multiplication operations that may introduce information bottlenecks or artifacts. However, compared to mask-based methods, the physical significance of mapping methods is often less intuitive, with their internal separation mechanism resembling more of a ``black box", reducing model interpretability. Moreover, learning direct mapping functions from mixed signals to multiple independent source signal representations may require more complex network structures and larger parameter counts, especially when processing highly overlapping or nonlinearly mixed signals, potentially leading to higher computational costs and training optimization challenges \cite{wang2023tf,dong2025edsep,chen2023sepdiff}. Therefore, although mapping methods theoretically provide a more direct optimization path, in practice, trade-offs must be made between model complexity, interpretability, and separation performance.

\section{Evaluation Metrics}
\label{sec:loss}
\begin{table*}[!t]
\centering
\caption{Comparison of subjective and objective metrics. ``Intru" denotes the metric is intrusive (i.e.\ requires a clean reference); ``Scale" denotes the metric's value range; ``Unit" denotes the unit or interpretation of the metric.}
\label{tab:metrics_compare}
\begin{tabular}{lllll}
\toprule
\multicolumn{5}{c}{\textbf{Subjective Metrics}}\\
\toprule
\textbf{Metric}                  & \textbf{Intru} & \textbf{Scale}   & \textbf{Unit}   & \textbf{Advantages/Limits}               \\
\midrule
MOS   & No        & 1--5    & Score  & Intuitive; high cost, time-consuming \\
\toprule
\multicolumn{5}{c}{\textbf{Objective Metrics}}\\
\toprule
\textbf{Metric}                  & \textbf{Intru} & \textbf{Scale}   & \textbf{Unit}   & \textbf{Advantages/Limits}               \\
\midrule
SDR\cite{vincent2006performance}      & Yes & dB & Ratio & Easy to compute; ignores semantics/intelligibility \\
SIR\cite{vincent2006performance}      & Yes & dB & Ratio & Interference suppression; no distortion measure   \\
SAR\cite{vincent2006performance}      & Yes & dB & Ratio & Artifact evaluation; ignores residual interference \\
SI\-SDR\cite{le2019sdr}               & Yes & dB & Ratio & Scale-invariant; robust to convolutive distortion  \\
SDRi\cite{vincent2006performance}     & Yes & dB & Gain  & Relative gain; no absolute quality measure         \\
SI\-SDRi\cite{le2019sdr}              & Yes & dB & Gain  & Relative + invariant; only measures improvement    \\
CI\-SDR\cite{vincent2006performance}  & Yes & dB & Ratio & Covers noise + artifacts; complex definition       \\
SA\-SDR\cite{von2022sa}               & Yes & dB & Ratio & Multi-channel; aggregation-dependent               \\
SA\-SI\-SDR\cite{von2022segment}      & Yes & dB & Ratio & Multi + invariant; computationally heavy           \\
SA\-CI\-SDR\cite{von2022segment}      & Yes & dB & Ratio & Multi + convolution-invariant; filter estimation hard \\
PESQ (NB)\cite{rix2001perceptual}     & Yes & –0.5$\sim$4.5 & MOS-like & Models auditory masking; standard narrowband range \\
PESQ (WB)\cite{rix2001perceptual}     & Yes & –0.5$\sim$4.5 & MOS-like & Supports wideband speech; improved fidelity        \\
STOI\cite{taal2011algorithm}          & Yes & 0$\sim$1      & Corr.    & Accurate intelligibility; noise-sensitive          \\
ESTOI\cite{jensen2016algorithm}       & Yes & 0$\sim$1      & Corr.    & Extended; better correlation under modulated noise \\
DNSMOS\cite{reddy2020interspeech}     & No  & 1$\sim$5      & Score    & Non-intrusive; needs large-scale annotations       \\
SIGMOS\cite{ristea2025icassp}         & No  & 1$\sim$5      & Score    & End-to-end; complex training/deployment            \\
\bottomrule
\end{tabular}
\label{tab:metrics}
\vspace{-15pt}
\end{table*}

Speech separation evaluation is a critical step for measuring performance, guiding algorithm optimization, and facilitating practical applications. With the widespread application of speech separation technology in communication \cite{divenyi2004speech}, human-computer interaction \cite{fadil2024review}, hearing assistance \cite{wang2008time}, and other fields, establishing a scientific, comprehensive, and reliable evaluation system has become particularly important. An ideal speech separation system should accurately separate target speech while maintaining speech naturalness and intelligibility, and minimizing interference, noise, and processing-induced distortion. Evaluation metrics are generally classified into two categories: subjective evaluation metrics and objective evaluation metrics. Subjective evaluation relies on human listeners' direct perceptual assessment of separation results, reflecting the actual perceptual quality of speech separation. Objective evaluation, on the other hand, provides quantitative analysis of separation results through mathematical models and algorithms, offering reproducible and efficient assessment methods. As shown in Table \ref{tab:metrics}, these two evaluation approaches have their respective advantages, complement each other, and together constitute a complete system for evaluating speech separation technology.

\subsection{Subjective Metrics}
Subjective evaluation methods primarily rely on human listeners' direct judgment of speech quality, with the Mean Opinion Score (MOS) rating system being the most widely used. In MOS evaluation, listeners typically score speech samples on a scale of 1-5 according to the ITU-T P.800 standard \cite{union2001international}, where 1 indicates extremely poor quality and 5 indicates excellent quality. MOS ratings can be subdivided into multiple dimensions, such as Speech Quality MOS (SIG), Background Noise MOS (BAK), and Overall MOS (OVRL) \cite{hu2007evaluation}. Additionally, the ITU-T P.835 standard \cite{itu2003835} is specifically designed for evaluating the performance of noise suppression algorithms, while the MUSHRA (MUltiple Stimuli with Hidden Reference and Anchor) test \cite{series2014method} is used for assessing high-quality audio systems. The primary advantage of subjective evaluation is that it directly reflects end-users' perceptual experience, making the evaluation results highly valuable for practical applications.

However, subjective evaluation also faces numerous challenges: first, organizing large-scale subjective tests requires substantial human resources and time; second, test results can be influenced by individual differences among listeners, testing environments, cultural backgrounds, and auditory fatigue, leading to subjective bias and consistency issues; third, subjective testing is difficult to automate and standardize, making it unsuitable for real-time optimization guidance during algorithm development. Nevertheless, subjective evaluation is still considered the ``golden standard" for assessing speech separation technology, and the effectiveness of objective evaluation metrics is typically validated through their correlation with subjective evaluation results.

\subsection{Objective Metrics}

Objective evaluation metrics quantitatively analyze the relationship between separated speech and reference signals through mathematical models, providing an automated, reproducible, and efficient assessment method. In signal-level evaluation, the Signal-to-Noise Ratio (SNR) is the most fundamental metric, used to measure the energy ratio between the target speech signal and background noise \cite{vincent2006performance}. The Signal-to-Distortion Ratio (SDR) and its improved versions offer a more comprehensive assessment, measuring the energy ratio between the reference signal and the estimated signal. The SDR improvement (SDRi) evaluates algorithm performance by comparing the changes in SDR before and after processing \cite{vincent2006performance}. To address the sensitivity of traditional SDR to signal scaling, the Scale-Invariant Signal-to-Distortion Ratio (SI-SDR) and the Scale-Invariant SDR improvement (SI-SDRi) have been proposed \cite{le2019sdr}, introducing an optimal scaling factor to ensure fairer evaluations. In multi-source separation scenarios, the Convolution-Invariant Signal-to-Distortion Ratio (CI-SDR) considers various error terms, including interference, noise, and distortion introduced by the algorithm \cite{vincent2006performance}. For conference-type multi-talker speech data, the Source-Aggregated Signal-to-Distortion Ratio (SA-SDR) \cite{von2022sa} evaluates separation performance by aggregating multi-channel signals, with derived metrics SA-SI-SDR and SA-CI-SDR \cite{von2022segment} targeting scale-invariant and convolution-invariant applications, respectively. Additionally, the Signal-to-Interference Ratio (SIR) and the Signal-to-Artifacts Ratio (SAR) \cite{vincent2006performance} measure the extent of target signal interference suppression and algorithm-induced distortion, often used in conjunction with SDR to form a more comprehensive evaluation.

On the perceptual level, the Perceptual Evaluation of Speech Quality (PESQ) \cite{rix2001perceptual} assesses speech quality by simulating the human auditory system, while the Short-Time Objective Intelligibility (STOI) \cite{taal2011algorithm} predicts speech intelligibility by calculating short-term correlation. The Perceptual Objective Listening Quality Assessment (POLQA) \cite{beerends2013perceptual}, as a third-generation speech quality assessment algorithm, supports wideband speech evaluation and accurately measures nonlinear distortion, time-varying distortion, and time-scaling distortion. With the advancement of deep learning technologies, no-reference metrics such as the Deep Noise Suppressor Mean Opinion Score (DNSMOS) \cite{reddy2020interspeech} and SigMOS \cite{ristea2025icassp} directly predict subjective ratings through neural network analysis without requiring reference signals, demonstrating broad prospects in practical applications. The main advantages of objective evaluation metrics lie in their automation, low cost, and high reproducibility; however, they also have certain limitations: many objective metrics struggle to fully capture the complexity of human auditory perception, particularly regarding subjective experiences such as speech naturalness and auditory comfort; different metrics are designed for different application scenarios, and no single metric is suitable for all speech separation tasks; some metrics (e.g., PESQ-NB) are primarily designed for narrowband speech and may not be accurate enough in wideband speech or music scenarios.

While objective metrics are automated, low-cost, and reproducible, they have limitations. Many struggle to fully capture the complexity of human auditory perception, such as naturalness and comfort. Furthermore, the sheer number of metrics can make direct comparison across different studies difficult if evaluation standards are not shared. To address this and promote consistency, it is highly recommended that studies report a core set of metrics that provide a multi-faceted view of performance. For a comprehensive evaluation, the standard practice should include SI-SDR (or SI-SDRi) as the primary metric for separation accuracy, alongside traditional SDR for backward compatibility. To assess perceptual impact, reporting PESQ for speech quality and STOI (or its successor, ESTOI) for speech intelligibility is essential. Reporting this combination of metrics (SI-SDR, SDR, PESQ, and STOI/ESTOI) provides a balanced assessment covering both signal fidelity and human perception, enabling more meaningful and reliable comparisons across different algorithms.

\section{Datasets}
\label{sec:dataset}
\begin{table}[!t]
\scriptsize
\centering
\caption{A comparative overview of monaural and multi-channel datasets categorized by acoustic characteristics and recording conditions. The table highlights key parameters including noise conditions, reverberation properties, speaker overlap patterns, and licenses.}
\begin{tabular}{lllll}
\toprule
\textbf{Dataset} & \textbf{Noise} & \textbf{Reverb} & \textbf{Overlap} & \textbf{License} \\ \midrule
\multicolumn{5}{c}{\textbf{Monaural}} \\ \midrule
WSJ0 \cite{hershey2016deep}            & \xmark & \xmark & 100\%   & LDC          \\
WHAM! \cite{wichern2019wham}                & \cmark & \xmark & 100\%   & CC BY-NC 4.0 \\
WHAMR! \cite{maciejewski2020whamr}          & \cmark & \cmark & 100\%   & CC BY-NC 4.0 \\
LibriMix \cite{cosentino2020librimix}       & \cmark & \xmark & 0–100\% & MIT          \\
DNS Challenge \cite{reddy2021interspeech}   & \cmark & \cmark & 100\%   & MIT          \\ 
REAL-M \cite{subakan2022real}               & \cmark & \cmark & 100\%   & Apache-2.0   \\
Lombard-GRID \cite{ewert2024does}           & \cmark & \xmark & 100\%   & Apache-2.0   \\
LibriheavyMix \cite{jin2024libriheavymix}   & \cmark & \cmark & 0–100\% & Apache-2.0   \\
LRS2-2Mix \cite{li2024audio}                & \cmark & \cmark & 100\%   & Academic     \\
SonicSet \cite{li2024sonicsim}              & \cmark & \cmark & 0–100\% & CC BY-SA 4.0 \\ \midrule
\multicolumn{5}{c}{\textbf{Multi-channel}} \\ \midrule
SMS-WSJ \cite{drude2019sms}                 & \cmark & \cmark & 100\%   & LDC          \\
LibriCSS \cite{chen2020continuous}          & \cmark & \cmark & 0–40\%  & MIT          \\
Kinect-WSJ \cite{sivasankaran2021analyzing} & \cmark & \cmark & 0–100\% & LDC          \\
AISHELL-4 \cite{fu2021aishell}              & \cmark & \cmark & 0–100\% & CC BY-SA 4.0 \\ \bottomrule
\end{tabular}
\label{tab:datasets}
\vspace{-10pt}
\end{table}
Speech separation, as a crucial task in the field of speech processing, heavily relies on the availability of high-quality training data. Most speech separation methods employ supervised learning, which requires that the training data provide both clean source signals and mixture signals, enabling the model to effectively learn to separate mixed speech into individual speaker voices~\cite{hershey2016deep,subakan2022real}. Existing speech separation datasets can be classified into single-channel and multi-channel categories, as illustrated in Table~\ref{tab:datasets}. These datasets capture various aspects of real-world acoustic environments, including reverberation, device noise, speech overlap, and background interference.

\subsection{Monaural}

Single-channel speech separation tasks require high-quality multi-speaker mixed speech data, comprising both clean source signals and mixture signals, in order to enable models to learn to separate mixed speech into individual speaker voices. Existing datasets can be categorized into three types: (1) simple synthetic mixtures based on clean speech, such as WSJ0-2mix~\cite{hershey2016deep} and WSJ0-3mix; (2) complex mixtures incorporating environmental noise and reverberation, such as WHAM!~\cite{wichern2019wham} and WHAMR!~\cite{maciejewski2020whamr}; and (3) large-scale datasets simulating real-world scenarios, such as LibriMix~\cite{cosentino2020librimix} and LibriheavyMix~\cite{jin2024libriheavymix}. The early dataset WSJ0-2mix includes only 101 speakers and a limited vocabulary, resulting in inadequate model generalization. To address this limitation, LibriMix leverages the 1,252 speakers from LibriSpeech to construct a larger-scale mixed speech dataset, and additionally introduces sparse overlap patterns to better simulate real conversational scenarios. With the progression of research, dataset design has gradually shifted towards greater diversity. For instance, LibriheavyMix provides 20,000 hours of multi-speaker speech with accurate time-stamped annotations; LRS2-2Mix~\cite{li2024audio} is constructed based on BBC television programs, encompassing a variety of accents and environmental variations; SonicSet~\cite{li2024sonicsim} utilizes the SonicSim tool to simulate moving sound sources in 90 distinct scenarios, thereby enhancing spatial acoustic features; Lombard-GRID-2mix~\cite{ewert2024does} focuses on the effect of speaker articulation changes under noisy environments on speech separation performance. The latest trend is the development of real-recorded speech mixture datasets, such as REAL-M~\cite{subakan2022real},and DNS Challenge datasets \cite{reddy2021interspeech} tailored for noise suppression tasks. In the future, the development of datasets may focus on: (1) recording at larger scales and with greater diversity in real-world scenarios; (2) providing fine-grained annotations, including speaker emotion, accent, age, and other attributes; (3) constructing multimodal datasets by combining visual, textual, and other information to facilitate speech separation; and (4) building continuous speech separation datasets in dynamic scenarios to support the development of real-time application systems.

\subsubsection{WSJ0}

The Wall Street Journal (WSJ0) corpus is a high-quality, read-speech collection originally created for automatic speech recognition research \cite{garofolo1993darpa}. Due to its clean recordings, it has become the standard foundation for creating benchmark speech separation datasets. The most prominent of these is wsj0-2mix \cite{hershey2016deep}, which generates mixtures of two speakers by combining their signals at various signal-to-noise ratios. To address more complex scenarios, the wsj0-3mix dataset extends this by creating mixtures from three distinct speakers. Collectively, these datasets have been instrumental as standard benchmarks for developing and evaluating speech separation algorithms in multi-talker environments.

\subsubsection{WHAM! \& WHAMR!}
The WHAM! (Wall Street Journal 0 Habitat multi-channel) dataset extends the WSJ0 2-mix dataset by adding environmental noise from the real world to simulate speech separation problems in various noise scenarios \cite{wichern2019wham}. It includes different noise conditions representing indoor and outdoor environments, public spaces, and transportation vehicles, all sourced from real-world recordings. The WHAMR! dataset is an extension of the WHAM dataset, featuring not only background noise but also reverberation effects \cite{maciejewski2020whamr}. This provides data support for speech separation research under more complex acoustic conditions.

\subsubsection{LibriMix}
LibriMix \cite{cosentino2020librimix} is a foundational open-source speech separation dataset, created to overcome the significant generalization limitations of prior benchmarks like wsj0-2mix \cite{hershey2016deep}, which suffer from limited speaker and vocabulary diversity and unrealistic, fully-overlapped mixtures. To address this, LibriMix leverages the large-scale LibriSpeech corpus \cite{panayotov2015librispeech} and WHAM! noise \cite{wichern2019wham} to generate a much more diverse and challenging dataset. Its construction pipeline features a baseline version that mixes speakers using perceptually-aware LUFS normalization \cite{series2011algorithms} and adds environmental noise. Crucially, it also introduces a sparse version designed to mimic natural conversation, which uses a forced aligner \cite{mcauliffe2017montreal} to create longer mixtures with varying degrees of temporal overlap (0\%–100\%). This dual approach, offering both clean and noisy mixtures with dense and sparse overlaps, provides a comprehensive and scalable benchmark for training and evaluating more generalizable speech separation models.

\subsubsection{DNS Challenge}

The DNS (Deep Noise Suppression) Challenge dataset \cite{reddy2021interspeech} is a key resource in the field of speech separation and noise suppression. Developed to support the DNS Challenge hosted by Microsoft, this dataset includes a diverse collection of real and synthetic noises designed to mimic a variety of acoustic environments. It also contains clean speech recordings from multiple speakers in various languages, which are mixed with background noises to create challenging audio scenarios for testing and enhancing noise suppression algorithms. The DNS dataset is widely utilized by researchers and developers to train and evaluate their models, facilitating advancements in the domain of speech processing technologies by providing a rich and varied set of audio samples for comprehensive testing and development.

\subsubsection{REAL-M}
REAL-M \cite{subakan2022real} is a real-world mixed speech dataset collected through crowdsourcing, specifically designed for two-speaker scenarios. This dataset comprises over 1,436 mixed speech samples, sourced from 50 speakers, including both native and non-native English speakers, with a total duration of approximately 3 hours. The mixed speech samples are generated by participants synchronously reading paired texts randomly selected from the LibriSpeech test set, with sentence lengths strictly constrained between 5 and 15 words.  A distinguishing characteristic of the mixing process is the diversity of recording conditions: participants utilize various consumer-grade devices, such as laptops and smartphones, and perform recordings in different acoustic environments. These settings encompass near-field and far-field scenarios, as well as varying levels of environmental noise and reverberation effects. Each mixed audio segment is accompanied by precise text transcriptions, providing a crucial benchmark for subsequent automatic speech recognition (ASR) performance evaluation.

\subsubsection{Lombard-GRID-2mix}
The Lombard-GRID-2mix dataset \cite{ewert2024does} is specifically designed to investigate the impact of the Lombard effect—changes in articulation due to noisy environments—on speech separation systems. Derived from the Audio-Visual Lombard GRID corpus \cite{alghamdi2018corpus}, it comprises two parallel subsets containing two-speaker mixtures: one with normal speech and another with Lombard speech. Both subsets are constructed by mixing two speakers at an 8~kHz sampling rate with a relative signal-to-noise ratio between 0–5~dB, using a "min" clipping mode to ensure maximum temporal overlap. Critically, to simulate the conditions that induce the Lombard effect, the Lombard subset further incorporates speech-shaped background noise (SSN) into the mixtures. This dual-subset design, which includes meticulously partitioned training, validation, and test sets, provides a controlled environment to systematically analyze how speech style variations affect the performance of speech separation algorithms.

\subsubsection{LibriheavyMix}
LibriheavyMix \cite{jin2024libriheavymix} is a massive 20,000-hour, single-channel dataset of far-field overlapping speech, designed to overcome the data scarcity and lack of realism in prior corpora for complex multi-speaker tasks. Built upon the Libriheavy \cite{kang2024libriheavy} corpus, it employs a sophisticated synthesis pipeline: first, a dynamic overlap algorithm, informed by the statistical properties of real meetings \cite{yu2022m2met}, generates complex multi-turn dialogues for 1 to 4 speakers. Next, these dialogues are convolved with simulated room impulse responses using the FAST-RIR toolkit \cite{luo2022fra} to create diverse and realistic far-field reverberant conditions. Critically, the dataset provides precise, time-aligned annotations including speaker identity, text, and punctuation, establishing it as a foundational resource for developing and evaluating jointly optimized, end-to-end systems for speech separation, recognition, and speaker diarization.

\subsubsection{LRS2-2Mix}
The LRS2-2Mix dataset \cite{li2024audio} is a synthetic two-speaker mixed speech benchmark derived from the LRS2-BBC audio-visual corpus \cite{afouras2018deep}, which comprises real-world video clips from an extensive collection of BBC programs. It is created by mixing speech signals from different speakers found in the LRS2-BBC validation and test sets at signal-to-noise ratios (SNRs) ranging from -5 dB to 5 dB. The resulting dataset provides approximately 11 hours of training data, along with 3-hour validation and test sets. By retaining the associated visual information from the source, LRS2-2Mix serves as a valuable resource for developing and evaluating audio-visual speech separation and recognition models in multi-talker scenarios.

\subsubsection{SonicSet}
SonicSet \cite{li2024sonicsim} is a large-scale, high-quality dataset designed to provide realistic and diverse training data of moving sound sources for speech separation and enhancement tasks. It integrates 360 hours of speech from LibriSpeech \cite{panayotov2015librispeech} and various noises into 90 distinct 3D scenes sourced from Matterport3D \cite{chang2017matterport3d}. The dataset was created using the SonicSim synthesis tool \cite{li2024sonicsim}, which simulates the movement of sound sources by generating trajectories within the 3D environments. Along these paths, it computes time-varying Room Impulse Responses (RIRs) and convolves them with source audio to produce dynamic, spatialized recordings that closely mimic real-world acoustics. In addition to the audio, SonicSet provides rich metadata in JSON files detailing source trajectories and positions, making it an exceptionally valuable resource not only for separation and enhancement, but also for related tasks like sound source localization and voice activity detection.

\subsection{Multi-channel}

Multi-channel speech signal processing holds critical significance in real-world applications. Compared with single-channel methods, multi-channel methods can effectively exploit spatial information, thereby markedly enhancing speech separation performance. To advance research in this field, various multi-channel datasets with diverse characteristics have been constructed. Existing multi-channel speech separation datasets can be broadly categorized into two types: simulated reverberant datasets (e.g., SMS-WSJ~\cite{drude2019sms}) and datasets collected with real devices (e.g., Kinect-WSJ~\cite{sivasankaran2021analyzing}, AISHELL-4~\cite{fu2021aishell}, and LibriCSS~\cite{chen2020continuous}). These datasets capture different aspects of real-world acoustic environments. SMS-WSJ \cite{drude2019sms} creates multi-speaker environments with simulated reverberation by combining the WSJ0 corpus with simulated room acoustics; Kinect-WSJ \cite{sivasankaran2021analyzing} focuses on noise and spatial characteristics captured by real devices; AISHELL-4 records complex acoustic environments from genuine meeting scenarios, including short pauses, speech overlaps, and background noise; LibriCSS targets continuous speech separation scenes, providing speech data with varying overlap ratios. Although current datasets have significantly propelled research in multi-channel speech separation, future dataset development still faces several challenges and opportunities, such as constructing larger-scale and more diverse datasets, collecting data from more complex and dynamic scenarios (e.g., speaker motion and multiple noise sources), and exploring multimodal datasets that integrate auxiliary information such as vision and motion to assist speech separation. These directions will better reflect the complexity of practical application scenarios and further bridge the gap between laboratory research and real-world deployment.

\subsubsection{SMS-WSJ}

The SMS-WSJ dataset is a multi-channel speech separation and recognition dataset that combines audio signals from the WSJ0 corpus with simulated room acoustics\cite{drude2019sms}. In particular, SMS-WSJ leverages multi-channel room acoustic simulations to create a speech dataset that more closely resembles real-life scenarios. It is particularly well-suited for the development and evaluation of algorithms for speech separation and recognition in complex acoustic environments. The dataset includes not only the original clean speech signals but also reverberant speech signals simulated in a multi-talker environment, posing additional challenges for speech separation and recognition systems.

\subsubsection{LibriCSS}
The LibriCSS dataset \cite{chen2020continuous} is designed to evaluate Continuous Speech Separation (CSS) by simulating realistic conversational scenarios. It was created by concatenating speech segments from LibriSpeech \cite{panayotov2015librispeech} to form 10 hours of continuous audio, which was then re-recorded in a meeting room using a seven-channel circular microphone array to capture authentic room acoustics. The dataset is structured into one-hour sessions, each containing mini-sessions with controlled speaker overlap ratios varying from 0\% to 40\%. Each mini-session features eight speakers randomly selected from LibriSpeech's ``test clean'' subset, with speakers positioned at fixed but random locations. To enable systematic assessment, LibriCSS is equipped with a Kaldi-based \cite{povey2011kaldi} ASR evaluation protocol, allowing for the direct measurement of recognition performance on separated speech.

\subsubsection{Kinect-WSJ}

The Kinect-WSJ dataset \cite{sivasankaran2021analyzing} is a challenging audio dataset that consists of speech recordings from the WSJ0 corpus captured using the Microsoft Kinect device. The aim of this dataset is to facilitate speech recognition research in real-world noise conditions and potential room reverberation. The Kinect-WSJ dataset places a particular emphasis on the characteristics of the Kinect microphone array when capturing speech, which include higher noise levels and spatial audio capturing capabilities, making this dataset particularly relevant for automatic speech recognition and sound source localization studies.

\subsubsection{AISHELL-4}
The AISHELL-4 \cite{fu2021aishell} is a challenging dataset designed to bridge the gap between research and real-world meeting applications. It comprises 120 hours of audio from 211 real meetings, recorded with an 8-channel circular microphone array and featuring 4 to 8 native Mandarin speakers per session. The dataset captures complex acoustic environments, including speech overlaps, rapid speaker turns, and diverse background noises (e.g., keyboard typing, fan noise), to increase processing difficulty. To ensure realism, recordings were conducted in rooms of varying sizes and materials, with speaker-to-microphone distances ranging from 0.6m to 6.0m; speakers were static in the training set but mobile in the evaluation set. For supervised learning, the data is meticulously annotated with explicit markings of overlapping speech regions. All information, including speaker IDs, gender, precise timestamps, and transcriptions, is provided in the TextGrid format, making AISHELL-4 highly suitable for developing and evaluating end-to-end multi-speaker processing pipelines, from signal processing to ASR and diarization.

\section{Results with Different Models and Different Datasets}
\label{sec:results}
In recent years, the field of speech separation has witnessed rapid development, with improvements in model performance largely propelled by the advent of high-quality, large-scale benchmark datasets. The currently widely adopted standard datasets include WSJ0-2Mix~\cite{hershey2016deep}, WHAM!~\cite{wichern2019wham}, LibriMix~\cite{cosentino2020librimix}, and WHAMR!~\cite{maciejewski2020whamr}, which cover scenarios ranging from clean speech to complex acoustic conditions with intense noise and reverberation. Furthermore, newly proposed datasets such as LRS2-2Mix~\cite{xu2024tiger} and SonicSet~\cite{li2024sonicsim} present authentic challenges under realistic acoustic environments. We have compiled the performance of various approaches across several mainstream datasets, focusing particularly on their improvements in standard metrics such as SI-SNRi and SDRi (see Tables~\ref{tab:result-wsj0-2mix},~\ref{tab:result-wham},~\ref{tab:result-librimix},~\ref{tab:result-whamr},~\ref{tab:result-lrs2_2mix},~\ref{tab:result-sonicset}), and we also compare model efficiency in terms of parameter count. Most experimental results are directly reproduced from the original publications. For works that do not openly report specific scores, we adopt evaluation data from subsequent related studies to ensure fairness in horizontal comparison.

\begin{table}[!t]
\centering
\caption{Performance comparison of speech separation models on the WSJ0-2Mix dataset. The ``Type'' column denotes the model paradigm: ``G'' for generative and ``D'' for deterministic. Same conventions are used in subsequent tables.}
\begin{tabular}{l|cccc}
\toprule
\textbf{Model} & \textbf{SI-SDRi} & \textbf{SDRi} & \textbf{Params (M)} & \textbf{Type} \\
\midrule
SPMamba \cite{li2024spmamba}                     & 22.5 & 22.7 & 6.1   & D \\
EDSep \cite{dong2025edsep}                       & 15.9 & –    & –     & G \\
ReSepFormer \cite{della2024resource}             & 18.6 & 18.9 & 8.0   & D \\
SepTDA \cite{lee2024boosting}                    & 24.0 & 23.9 & 21.2  & D \\
MossFormer2 \cite{zhao2024mossformer2}           & 24.1 & –    & 55.7  & D \\
Separate And Diffuse \cite{lutati2023separate}   & 23.9 & –    & –     & G \\
S4M \cite{chen2023neural}                        & 20.5 & 20.7 & 3.6   & D \\
pSkiM \cite{li2023predictive}                    & 15.5 & –    & 8.5   & D \\
DiffSep \cite{scheibler2023diffusion}            & 14.3 & –    & –     & G \\
TF-GridNet \cite{wang2023tf}                     & 23.5 & 23.6 & 14.5  & D \\
MossFormer \cite{zhao2023mossformer}             & 22.8 & –    & 42.1  & D \\
TDANet \cite{li2022efficient}                    & 18.6 & 18.9 & 2.3   & D \\
SepEDA \cite{chetupalli2022speech}               & 21.2 & 21.4 & 12.5  & D \\
SkiM \cite{li2022skim}                           & 18.3 & 18.7 & 5.9   & D \\
MTDS \cite{qian2022efficient}                    & 21.5 & 21.7 & 4.0   & D \\
QDPN \cite{rixen2022qdpn}                        & 23.6 & –    & 200.0 & D \\
SFSRNet \cite{rixen2022sfsrnet}                  & 24.0 & 24.1 & 59.0  & D \\
TFPSNet \cite{yang2022tfpsnet}                   & 21.1 & 21.3 & 2.7   & D \\
Unknow-SS \cite{chazan2021single}                & 19.4 & –    & –     & D \\
A-FRCNN \cite{hu2021speech}                      & 18.3 & 18.6 & 6.1   & D \\
Sandglasset \cite{lam2021sandglasset}            & 20.8 & 21.0 & 2.3   & D \\
SepFormer \cite{subakan2021attention}            & 22.3 & 22.4 & 26.0  & D \\
WaveSplit \cite{zeghidour2020wavesplit}          & 22.3 & 22.4 & 29.0  & D \\
MSGT-TasNet \cite{zhao2021multi}                 & 17.0 & 17.3 & 66.8  & D \\
Multi-Decoder Dprnn \cite{zhu2021multi}          & 19.1 & –    & –     & D \\
DPTCN-ATPP \cite{zhu2021dptcn}                   & 19.6 & 19.9 & 4.7   & D \\
DPTNet \cite{chen2020dual}                       & 20.2 & 20.6 & 2.7   & D \\
Conv-TasSAN \cite{deng2020conv}                  & 15.1 & 15.4 & 5.0   & G \\
DPRNN \cite{luo2020dual}                         & 18.8 & 19.0 & 2.9   & D \\
VSUNOS \cite{nachmani2020voice}                  & 20.1 & 20.4 & 7.5   & D \\
Two-Step TCN \cite{tzinis2020two}                & 16.1 & –    & 8.6   & D \\
SudoRM-RF \cite{tzinis2020sudo}                  & 17.0 & 17.3 & 2.7   & D \\
Deep CASA \cite{liu2019divide}                   & 17.7 & 18.0 & 12.8  & D \\
Conv-TasNet \cite{luo2019conv}                    & 15.3 & 15.6 & 5.1   & D \\
OR-PIT \cite{takahashi2019recursive}             & 14.8 & 15.0 & –     & D \\
Chimera++ sign \cite{wang2019deep}               & 15.3 & 15.6 & –     & D \\
ADAN \cite{luo2017speaker}                       & 10.4 & 10.8 & 9.1   & D \\
TaSNet \cite{luo2018tasnet}                      & 13.2 & 13.6 & 23.6  & D \\
Chimera++ Network \cite{wang2018alternative}     & 11.5 & 12.0 & 32.9  & D \\
DANet \cite{chen2017deep}                        & 10.5 & –    & 9.1   & D \\
uPIT-BLSTM \cite{kolbaek2017multitalker}         & 9.8  & 10.0 & 92.7  & D \\
DPCL \cite{hershey2016deep}                      & 10.8 & –    & –     & D \\
\bottomrule
\end{tabular}
\label{tab:result-wsj0-2mix}

\end{table}

\begin{table}[!t]
\centering
\caption{Performance comparison of speech separation models on the WHAM! dataset.}
\begin{tabular}{l|cccc}
\toprule
\textbf{Model} & \textbf{SI-SDRi} & \textbf{SDRi} & \textbf{Params (M)} & \textbf{Type} \\
\midrule
SPMamba \cite{li2024spmamba}                 & 17.4 & 17.6 & 6.1  & D \\
ReSepFormer \cite{della2024resource}         & 14.1 & 14.4 & 8.0  & D \\
Fast-GeCo \cite{wang2024noise}               & 12.6 & –    & –    & G \\
MossFormer2 \cite{zhao2024mossformer2}       & 18.1 & –    & 55.7 & D \\
MossFormer \cite{zhao2023mossformer}         & 17.3 & –    & 42.1 & D \\
TDANet \cite{li2022efficient}                & 15.2 & 15.4 & 2.3  & D \\
Unknow-SS \cite{chazan2021single}            & 11.5 & –    & –    & D \\
A-FRCNN \cite{hu2021speech}                  & 14.5 & 14.8 & 6.1  & D \\
SepFormer \cite{subakan2021attention}        & 16.4 & –    & 26.0 & D \\
WaveSplit \cite{zeghidour2020wavesplit}      & 16.0 & 16.5 & 29.0 & D \\
MSGT-TasNet \cite{zhao2021multi}             & 13.1 & –    & 66.8 & D \\
DPTNet \cite{chen2020dual}                   & 14.9 & 15.3 & 2.7  & D \\
DPRNN \cite{luo2020dual}                     & 13.7 & 14.1 & 2.9  & D \\
VSUNOS \cite{nachmani2020voice}              & 15.2 & –    & 7.5  & D \\
SudoRM-RF \cite{tzinis2020sudo}              & 12.9 & 13.3 & 2.7  & D \\
Conv-TasNet \cite{luo2019conv}                & 12.7 & –    & 5.1  & D \\
TaSNet \cite{luo2018tasnet}                  & 12.0 & –    & 23.6 & D \\
Chimera++ Network \cite{wang2018alternative} & 10.0 & –    & 32.9 & D \\
\bottomrule
\end{tabular}
\label{tab:result-wham}

\end{table}

\begin{table}[!t]
\centering
\caption{Performance comparison of speech separation models on the WHAMR! dataset.}
\begin{tabular}{l|cccc}
\toprule
\textbf{Model} & \textbf{SI-SDRi} & \textbf{SDRi} & \textbf{Params (M)} & \textbf{Type} \\
\midrule
SPMamba \cite{li2024spmamba}            & 16.6 & 15.2 & 6.1   & D \\
MossFormer2 \cite{zhao2024mossformer2}  & 17.0 & –    & 55.7  & D \\
TF-GridNet \cite{wang2023tf}            & 17.3 & 15.8 & 14.5  & D \\
MossFormer \cite{zhao2023mossformer}    & 16.3 & –    & 42.1  & D \\
QDPN \cite{rixen2022qdpn}               & 14.4 & –    & 200.0 & D \\
SepFormer \cite{subakan2021attention}   & 14.0 & 13.0 & 26.0  & D \\
WaveSplit \cite{zeghidour2020wavesplit} & 13.2 & 12.2 & 29.0  & D \\
DPTNet \cite{chen2020dual}              & 11.2 & 10.6 & 2.7   & D \\
DPRNN \cite{luo2020dual}                & 10.3 & –    & 2.9   & D \\
VSUNOS \cite{nachmani2020voice}         & 12.2 & –    & 7.5   & D \\
SudoRM-RF \cite{tzinis2020sudo}         & 13.5 & –    & 2.7   & D \\
Conv-TasNet \cite{luo2019conv}           & 8.3  & –    & 5.1   & D \\
TaSNet \cite{luo2018tasnet}             & 10.9 & –    & 23.6  & D \\
\bottomrule
\end{tabular}
\label{tab:result-whamr}
\end{table}

\begin{table}[!t]
\centering
\caption{Performance comparison of speech separation models on the LibriMix (train-100 min) dataset.}
\begin{tabular}{l|cccc}
\toprule
\textbf{Model} & \textbf{SI-SDRi} & \textbf{SDRi} & \textbf{Params (M)} & \textbf{Type} \\
\midrule
SPMamba \cite{li2024spmamba}                   & 19.9 & 20.4 & 6.1  & D \\
TIGER \cite{xu2024tiger}                       & 18.0 & 18.3 & 0.8  & D \\
Conv-TasNet GAN \cite{lakandri2024exploring}   & 12.2 & 12.6 & –    & G \\
Fast-GeCo \cite{wang2024noise}                 & 13.0 & –    & –    & G \\
DIP \cite{wang2024speech}                      & 15.8 & 16.1 & –    & D \\
MossFormer2 \cite{zhao2024mossformer2}         & 21.7 & –    & 55.7 & D \\
Separate And Diffuse \cite{lutati2023separate} & 21.5 & –    & –    & G \\
S4M \cite{chen2023neural}                      & 16.9 & 17.4 & 3.6  & D \\
DiffSep \cite{scheibler2023diffusion}          & 9.6  & –    & –    & G \\
TF-GridNet \cite{wang2023tf}                   & 19.2 & 19.6 & 14.5 & D \\
MossFormer \cite{zhao2023mossformer}           & 19.7 & –    & 42.1 & D \\
TDANet \cite{li2022efficient}                  & 17.4 & 17.9 & 2.3  & D \\
SSL-SS \cite{huang2022investigating}           & 11.1 & –    & –    & D \\
SFSRNet \cite{rixen2022sfsrnet}                & 16.4 & 16.9 & 59.0 & D \\
TFPSNet \cite{yang2022tfpsnet}                 & 19.7 & 19.9 & 2.7  & D \\
A-FRCNN \cite{hu2021speech}                    & 16.7 & 17.2 & 6.1  & D \\
SepFormer \cite{subakan2021attention}          & 16.5 & 17.0 & 26.0 & D \\
WaveSplit \cite{zeghidour2020wavesplit}        & 15.8 & 15.9 & 29.0 & D \\
DPTNet \cite{chen2020dual}                     & 16.7 & 17.1 & 2.7  & D \\
DPRNN \cite{luo2020dual}                       & 14.1 & 14.6 & 2.9  & D \\
Two-Step TCN \cite{tzinis2020two}              & 12.0 & 12.5 & 8.6  & D \\
SudoRM-RF \cite{tzinis2020sudo}                & 13.5 & 14.0 & 2.7  & D \\
Conv-TasNet \cite{luo2019conv}                  & 12.2 & 12.7 & 5.1  & D \\
TaSNet \cite{luo2018tasnet}                    & 7.9  & 8.7  & 23.6 & D \\
Chimera++ Network \cite{wang2018alternative}   & 6.3  & 7.0  & 32.9 & D \\
uPIT-BLSTM \cite{kolbaek2017multitalker}       & 7.6  & 8.2  & 92.7 & D \\
\bottomrule
\end{tabular}
\label{tab:result-librimix}

\end{table}

For the scenario of clean speech separation, WSJ0-2Mix~\cite{hershey2016deep} and LibriMix \cite{cosentino2020librimix} are the most representative datasets. Recent dual-path network architectures (such as SepTDA~\cite{lee2024boosting}, SFSRNet~\cite{rixen2022sfsrnet}, TF-GridNet~\cite{wang2023tf}, and MossFormer2~\cite{zhao2024mossformer2}, etc.) generally attain SI-SDRi values of 20 dB or higher on WSJ0-2Mix, with some methods (such as SepTDA and SFSRNet) even surpassing 24 dB, which is a substantial improvement over earlier approaches like Conv-TasNet~\cite{luo2019conv} and DPRNN~\cite{luo2020dual}. A similar trend is observed on the LibriMix (train-100) subset, where methods such as MossFormer2~\cite{zhao2024mossformer2}, Separate And Diffuse~\cite{lutati2023separate}, TF-GridNet~\cite{wang2023tf}, and TFPSNet~\cite{yang2022tfpsnet} all achieve remarkable results, with SI-SDRi values typically ranging from 18 to 21 dB, outperforming conventional methods by a significant margin. It is noteworthy that on datasets with broader speaker coverage such as LibriMix, parameter-efficient networks (e.g., TIGER~\cite{xu2024tiger}, TDANet~\cite{li2022efficient}) can still deliver competitive performance with extremely low model complexity (for instance, TIGER comprises just 0.8M parameters), thereby demonstrating the dual advantages of U-Net structures in generalizability and efficiency.

\begin{table}[!t]
\centering
\caption{Performance comparison of speech separation models on the LRS2-2Mix dataset.}
\begin{tabular}{l|cccc}
\toprule
\textbf{Model} & \textbf{SI-SDRi} & \textbf{SDRi} & \textbf{Params (M)} & \textbf{Type} \\
\midrule
EDSep \cite{dong2025edsep}            & 9.6  & –    & –    & G \\
DIP \cite{wang2024speech}             & 12.0 & 12.4 & –    & D \\
TIGER \cite{xu2024tiger}              & 15.1 & 15.3 & 0.8  & D \\
S4M \cite{chen2023neural}             & 15.3 & 15.5 & 3.6  & D \\
TDANet \cite{li2022efficient}         & 14.2 & 14.9 & 2.3  & D \\
A-FRCNN \cite{hu2021speech}           & 13.0 & 13.3 & 6.1  & D \\
SepFormer \cite{subakan2021attention} & 13.5 & 13.8 & 26.0 & D \\
DPTNet \cite{chen2020dual}            & 13.3 & 13.6 & 2.7  & D \\
DPRNN \cite{luo2020dual}              & 12.7 & 13.0 & 2.9  & D \\
SudoRM-RF \cite{tzinis2020sudo}       & 11.0 & 11.4 & 2.7  & D \\
Conv-TasNet \cite{luo2019conv}         & 10.6 & 11.0 & 5.1  & D \\
\bottomrule
\end{tabular}
\label{tab:result-lrs2_2mix}
\vspace{-10pt}
\end{table}

\begin{table}[!t]
\centering
\caption{Performance comparison of speech separation models on the SonicSet dataset.}
\begin{tabular}{l|cccc}
\toprule
\textbf{Model} & \textbf{SI-SDRi} & \textbf{SDRi} & \textbf{Params (M)} & \textbf{Type} \\
\midrule
TF-GridNet \cite{wang2023tf}         & 15.4 & 16.8 & 14.5 & D \\
MossFormer \cite{zhao2023mossformer} & 14.7 & 16.0 & 42.1 & D \\
TDANet \cite{li2022efficient}        & 9.3  & 11.0 & 2.3  & D \\
A-FRCNN \cite{hu2021speech}          & 9.2  & 10.6 & 6.1  & D \\
DPRNN \cite{luo2020dual}             & 4.9  & 6.7  & 2.9  & D \\
SudoRM-RF \cite{tzinis2020sudo}      & 8.0  & 9.7  & 2.7  & D \\
Conv-TasNet \cite{luo2019conv}        & 4.8  & 7.1  & 5.1  & D \\
\bottomrule
\end{tabular}
\label{tab:result-sonicset}

\end{table}

More challenging datasets, such as WHAM!~\cite{wichern2019wham}, WHAMR!~\cite{maciejewski2020whamr}, LRS2-2Mix~\cite{xu2024tiger}, and SonicSet~\cite{wang2023tf}, impose stricter demands on models due to the incorporation of real noise, reverberation, far-field, and other adverse conditions. On WHAM! and WHAMR!, the latest models including MossFormer2~\cite{zhao2024mossformer2}, SPMamba~\cite{li2024spmamba}, and TF-GridNet~\cite{wang2023tf}, continue to maintain notable superiority. For instance, MossFormer2 and TF-GridNet achieve SI-SDRi scores of 17.0 and 17.3 dB on WHAMR!, respectively, which significantly surpasses the 8.3 dB achieved by Conv-TasNet~\cite{luo2019conv}, thus illustrating outstanding robustness. The separation difficulty is further elevated in LRS2-2Mix and SonicSet, where the SI-SDRi scores of mainstream models decline substantially. For example, TDANet~\cite{li2022efficient} and A-FRCNN~\cite{hu2021speech} obtain only 9–11 dB on SonicSet, while earlier methods typically fall below 7 dB. It is worth noting that some models, such as TF-GridNet~\cite{wang2023tf} and MossFormer~\cite{zhao2023mossformer}, still maintain leading performance on SonicSet, attesting to the generalization capability of next-generation architectures under extreme noise conditions. Despite the improved performance and generalization of the latest approaches in complex real-world scenarios compared to traditional architectures, there remains substantial room for progress under highly diverse conditions, which poses ongoing challenges for future research.

\section{Platforms}
\label{sec:platform}
\begin{table*}[!t]
\centering
\caption{Comparison of open-source speech separation toolkits.}
\setlength{\tabcolsep}{4pt}  
\begin{tabular}{l l l p{3cm} p{3cm} p{3cm} l}
\toprule
\textbf{Name} & \textbf{PL} & \textbf{DL} & \textbf{Datasets} & \textbf{Pre-trained Models} & \textbf{Deployment} & \textbf{License} \\
\midrule
nussl \cite{manilow2018northwestern} &
Python & PyTorch &
MUSDB18, WSJ0-2Mix, WHAM!, FUSS &
DPCL, Chimera, OpenUnmix, TasNet, DPRNN &
Python API &
MIT \\ \midrule

ONSSEN \cite{ni2019onssen} &
Python & PyTorch &
WSJ0-2Mix, DAPS, Edinburgh-TTS &
None &
Python API &
GPL-3.0 \\ \midrule

ESPNet-SE \cite{li2021espnet,lu2022espnet} &
Python & PyTorch &
SLURP-S, LT-S, WSJ0-2Mix &
DANet, Conv-TasNet, DPRNN, SVoice, DPTNet, SkiM, TF-GridNet &
Python API &
Apache-2.0 \\ \midrule

Asteroid \cite{pariente2020asteroid} &
Python & PyTorch &
WSJ0-2Mix, WHAM!, LibriMix, FUSS &
Conv-TasNet, DPRNN &
Python API &
MIT \\ \midrule

SpeechBrain \cite{speechbrain,speechbrainV1} &
Python & PyTorch &
LibriMix, WSJ0-2Mix, WHAM! &
SepFormer, Conv-TasNet &
Python API &
Apache-2.0 \\ \midrule

ClearerVoice-Studio \cite{ClearerVoice-Studio} &
Python & PyTorch &
SLURP-S, LT-S &
MossFormer2 &
Python API, JIT, ONNX &
Apache-2.0 \\ \midrule

WeSep \cite{wang24fa_interspeech} &
Python, C++ & PyTorch &
Libri2Mix &
Spex+ (Conv-TasNet), pBSRNN, pDPCCN, TF-GridNet (JIT/ONNX) &
Python API, JIT, ONNX, C++ &
Apache-2.0 \\ 
\bottomrule
\end{tabular}
\label{tab:toolkits}
\vspace{-15pt}
\end{table*}

Open-source toolkits in the field of speech separation have greatly accelerated the advancement and application of related research. By providing unified frameworks, standardized evaluation metrics, and convenient dataset interfaces, these toolkits have significantly lowered the barriers to entry and improved the reproducibility of experiments, as illustrated in Table~\ref{tab:toolkits}. Representative tools, such as \texttt{nussl}~\cite{manilow2018northwestern}, ONSSEN~\cite{ni2019onssen}, ESPNet-SE~\cite{li2021espnet,lu2022espnet}, Asteroid~\cite{pariente2020asteroid}, SpeechBrain~\cite{speechbrain,speechbrainV1}, ClearerVoice~\cite{ClearerVoice-Studio}, and WeSep~\cite{wang24fa_interspeech}, are primarily developed in Python and predominantly adopt PyTorch~\cite{paszke2019pytorch} as the deep learning backend, reflecting the widespread dominance of this framework within the community. A common feature among these toolkits is their modular design, which enables researchers to conveniently replace or extend components such as data processing, model architectures, and loss functions according to specific research requirements. Furthermore, the provision of pretrained models and standardized evaluation pipelines—such as the integration of BSS-Eval~\cite{vincent2006performance}, \texttt{mir\_eval}~\cite{raffel2014mir_eval}, or SI-SNR—is another shared characteristic, facilitating rapid reproduction of baseline results and enabling fair comparisons.

Despite these similarities, toolkits still vary distinctly in design philosophy, core functionalities, and application focus. \texttt{nussl}~\cite{manilow2018northwestern}, an early and comprehensive library, covers a range of techniques from traditional methods (e.g., mask estimation, matrix factorization) to deep learning approaches, with particular emphasis on standardized evaluation and data management. ONSSEN~\cite{ni2019onssen} is dedicated to deep-learning-based speech separation and enhancement, employing a modular design that utilizes \texttt{Librosa} and \texttt{Numpy} for feature processing. ESPNet-SE~\cite{li2021espnet,lu2022espnet} distinguishes itself by seamlessly integrating front-end speech processing (separation and enhancement) with downstream tasks such as ASR, supporting multi-task joint training and focusing on multichannel and far-field scenarios. Asteroid~\cite{pariente2020asteroid} is recognized for its comprehensive functionality and retention of native PyTorch features, introducing Kaldi-style recipes to enhance reproducibility; it is primarily oriented toward monaural tasks. SpeechBrain~\cite{speechbrain,speechbrainV1}, as a general-purpose speech processing toolkit, includes speech separation as one of its many supported tasks and introduces innovative training strategies such as dynamic mixing to enhance model generalization. ClearerVoice~\cite{ClearerVoice-Studio} places greater emphasis on practical applications, providing a unified inference platform and supporting audio-visual target speaker extraction. WeSep~\cite{wang24fa_interspeech} is specialized for target speaker extraction (TSE), integrating advanced speaker modeling techniques (such as combination with WeSpeaker~\cite{wang2023wespeaker}), employing online data simulation strategies to improve generalization to unseen data, and offering model export (JIT/ONNX) functionalities for efficient deployment.

The collective efforts represented by these toolkits have driven considerable progress in speech separation technologies. Nonetheless, future toolkits are expected to further enhance model robustness and generalization under realistic and complex acoustic conditions (e.g., reverberation, noise, far-field), to develop more efficient and lightweight models suitable for edge computing, to strengthen the utilization of multimodal information (such as audio-visual integration, as explored by ClearerVoice~\cite{ClearerVoice-Studio} and WeSep~\cite{wang24fa_interspeech}), and to deliver more user-friendly deployment solutions and interfaces. Additionally, future developments may expand to broader audio processing tasks, such as universal source separation or deeper joint optimization with higher-level speech understanding in end-to-end frameworks. A detailed introduction to these toolkits is provided below.

\subsection{nussl}

The Northwestern University Source Separation Library (\texttt{nussl}) \cite{manilow2018northwestern} is designed to provide a unified and extensible research platform for source separation. Implemented in Python, \texttt{nussl} adopts an object-oriented design paradigm, ensuring high scalability and ease of extension. This library integrates various source separation techniques, including mask-based estimation methods, matrix decomposition approaches, and deep learning-based models. Additionally, it offers a comprehensive suite of standardized evaluation tools, enabling researchers to compare algorithms within a consistent framework. 



\subsection{ONSSEN}
ONSSEN \cite{ni2019onssen} is an open-source library specifically designed for speech separation and enhancement tasks, aiming to provide a flexible and extensible platform that facilitates researchers in implementing state-of-the-art deep learning algorithms and conducting fair comparisons. ONSSEN leverages \texttt{Librosa} and \texttt{Numpy} for feature extraction and employs \texttt{PyTorch} as the backend for model training.  ONSSEN adopts a modular design, consisting of a data module, a neural network module, and a loss computation module. The data module is responsible for feature extraction and data loading, the neural network module implements various speech separation algorithms, and the loss computation module defines different loss functions. This modular structure enables researchers to effortlessly replace or extend individual components to accommodate diverse experimental requirements.

\subsection{ESPNet-SE}

ESPNet-SE \cite{li2021espnet,lu2022espnet} is a toolkit specifically designed for speech enhancement and speech separation, while also integrating downstream tasks such as automatic speech recognition (ASR). ESPNet-SE provides a comprehensive suite of both single-channel and multi-channel speech enhancement and separation methods, enabling joint training with tasks such as ASR, speech translation, and spoken language understanding. The framework supports multi-task learning strategies, allowing for the joint optimization of different training objectives, such as signal-level loss and ASR loss, thereby improving overall system performance. Furthermore, ESPNet-SE introduces two novel multi-channel datasets, SLURP-S and LT-S, to facilitate research on far-field speech understanding tasks. The potential applications of enhancement front-ends in these tasks are also demonstrated, highlighting their value in improving system robustness.

\subsection{Asteroid}
Asteroid \cite{pariente2020asteroid} is an open-source audio source separation toolkit based on PyTorch \cite{paszke2019pytorch}, specifically designed for researchers and practitioners. It aims to provide both robust functionality and ease of use to support research and applications in monaural audio source separation and speech enhancement tasks. The development of this toolkit is inspired by state-of-the-art neural network-based source separation systems and integrates a complete workflow encompassing data preparation, model training, and evaluation. 

Compared to existing toolkits such as \texttt{nussl} \cite{manilow2018northwestern}, ONSSEN \cite{ni2019onssen}, and Open-Unmix \cite{stoter2019open}, Asteroid distinguishes itself through its comprehensiveness and flexibility. It not only supports a wide range of datasets and diverse network architectures but also reproduces key results from important research papers using Kaldi-style recipes \cite{povey2011kaldi}. This significantly enhances research reproducibility and facilitates experimental workflows. The design philosophy of Asteroid emphasizes maintaining a balance between abstraction and usability, preserving the native characteristics of PyTorch code whenever possible. Additionally, it enables seamless integration of third-party code with minimal modifications while ensuring experiment reproducibility through command-line configuration.

\subsection{SpeechBrain}
SpeechBrain \cite{speechbrain,speechbrainV1} is an open-source and general-purpose speech processing toolkit implemented in PyTorch. It is designed to provide a concise, flexible, and efficient framework that supports multiple speech-related tasks, including speech recognition, speech enhancement, speaker recognition, and speech separation. This toolkit not only offers a wide range of pre-trained models and experimental recipes but also possesses strong extensibility, enabling researchers to rapidly construct, compare, and optimize various speech processing systems. The core architecture of SpeechBrain is built around a modular design, allowing seamless integration of models across different tasks and thereby facilitating end-to-end speech processing workflows.

SpeechBrain implements a variety of state-of-the-art speech separation models and supports training and evaluation on different speech corpora. It incorporates the dynamic mixing training strategy \cite{subakan2021attention}, which dynamically generates mixed speech samples during training rather than relying on a fixed training dataset. This approach effectively enhances the model's generalization capability and mitigates the risk of overfitting to specific data distributions. The speech separation module in SpeechBrain is fully implemented in PyTorch and supports end-to-end training, allowing researchers to explore methodologies such as joint training and transfer learning.

\subsection{ClearerVoice}
ClearVoice-Studio \cite{ClearerVoice-Studio} is a unified inference platform that integrates speech enhancement, speech separation, and audio-visual target speaker extraction. It is designed to streamline the application of pre-trained models in speech processing tasks and facilitate their integration into real-world projects. The platform provides a collection of pre-trained models covering various speech processing tasks and supports automatic model loading for inference, eliminating the need for manual downloads.

In the speech separation module of ClearVoice-Studio, the system incorporates the MossFormer2\_SS\_16K model \cite{zhao2024mossformer2}, which is specifically optimized for speech data sampled at 16 kHz. Additionally, ClearVoice-Studio offers a flexible inference interface, supporting batch processing modes for single audio files, audio directories, and audio list files (.scp). Users can invoke MossFormer2\_SS\_16K for speech separation tasks through simple Python code and have the option to save the separated speech files online. This functionality provides researchers, engineers, and developers with a convenient experimental environment, enabling seamless integration of the model into various speech processing pipelines and application systems.

\subsection{Wesep}

WeSep \cite{wang24fa_interspeech} is an open-source toolkit designed for Target Speaker Extraction (TSE), providing an efficient, flexible, and scalable solution for both academic research and real-world applications. The TSE task focuses on accurately extracting the speech signal of a specific target speaker from a mixture of speech signals. It finds widespread applications in personalized human-computer interaction, hearing aids, as well as downstream tasks such as automatic speech recognition and speaker recognition. However, current TSE research faces challenges such as poor generalization to unseen data and a lack of efficient target speaker modeling methods, with relatively few open-source toolkits available. To bridge this gap, WeSep integrates mainstream TSE models, introduces advanced speaker modeling techniques, and offers efficient data management and online data simulation to facilitate advancements in TSE research.

\section{Challenges \& Explorations}
\label{sec:applicationtopics}
\subsection{Long-Form Audio Processing}



Long-form audio processing presents multiple challenges in the field of speech separation, significantly constraining the practical applicability of existing models. When handling long sequential inputs, speech separation models often encounter computational resource bottlenecks \cite{subakan2021attention,wang2023tf,chen2020dual,rixen2022sfsrnet}. This issue is particularly pronounced in Transformer-based architectures that employ self-attention mechanisms, such as SepFormer \cite{subakan2021attention} and TF-GridNet \cite{wang2023tf}, whose computational complexity scales quadratically with input length. Consequently, these models suffer from excessive computational overhead when processing long-duration audio. Moreover, many speech separation techniques rely on fixed-length windowing approaches to mitigate computational demands. However, such methods frequently result in the loss of crucial information at window boundaries, leading to degraded separation performance \cite{wang2023tf}. In particular, when speech content spans across window boundaries, models may struggle to correctly identify and separate speaker characteristics within continuous speech segments. 

Another critical challenge in long-sequence processing is the degradation of long-range dependency modeling capabilities. As the audio length increases, the model's ability to capture temporal dependencies across distant time points deteriorates \cite{chen2020continuous}. This limitation is particularly detrimental in accurately separating speech from different speakers, especially in scenarios with frequent speaker alternations or similar vocal characteristics. In real-world applications, such as meeting transcription, telemedicine consultations, and multi-speaker conversation analysis, speech content often spans extended durations. The limitations of conventional models severely affect the quality and usability of the separation results in such cases.

To address these challenges, researchers have proposed various solutions. Dual-path processing architectures, such as DPRNN \cite{luo2020dual} and DPTNet \cite{chen2020dual}, alternate between intra-chunk and inter-chunk processing, effectively balancing local feature extraction and global context modeling. The MSGT-TasNet \cite{zhao2021multi} incorporates a hierarchical attention mechanism that applies attention computation at different abstraction levels, thereby reducing overall computational complexity while preserving long-range modeling capabilities. Additionally, MTDS \cite{qian2022efficient} leverages a sliding window approach combined with an overlap-add strategy, widely adopted for processing arbitrary-length audio while alleviating boundary effects through inter-window information sharing. More advanced solutions include sparse attention mechanisms, such as linear attention \cite{shen2021efficient,katharopoulos2020transformers,wang2020linformer}, local attention \cite{mirsamadi2017automatic,chen2021regionvit}, and axial attention \cite{ho2019axial,valanarasu2021medical}, which restrict the scope of attention computation or modify its operation, reducing complexity from $\mathcal{O}(n^2)$ to $\mathcal{O}(n \log n)$ or even $\mathcal{O}(n)$. The MossFormer series \cite{zhao2023mossformer,zhao2024mossformer2} employs a hybrid local-global self-attention strategy, significantly alleviating computational burdens while maintaining global modeling capabilities. Furthermore, hybrid architectures that combine recurrent neural networks (RNNs) and convolutional neural networks (CNNs) have demonstrated advantages in long-sequence processing. For instance, models such as SkiM \cite{li2022skim} and pSkiM \cite{li2023predictive} leverage a skip memory mechanism to efficiently handle long-duration sequences.

\subsection{Lightweight Models}


Deploying lightweight models for speech separation remains a significant challenge, particularly in embedded systems, mobile applications, and edge computing platforms. Many high-performance models \cite{wang2023tf,zhao2023mossformer,zhao2024mossformer2,chen2020dual,subakan2021attention,lam2021sandglasset} are typically characterized by a large number of parameters and complex network architectures, resulting in high computational complexity and substantial memory consumption, which hinder deployment on resource-constrained devices. While model compression techniques such as pruning \cite{stamenovic2021weight} and quantization \cite{xu2022mixed,wu2023light} can mitigate computational demands, excessive compression may degrade separation performance.  Efforts to optimize attention module complexity are prevalent. For instance, S4M \cite{chen2023neural} and SPMamba \cite{li2024spmamba} leverage linear time-invariant systems for sequence modeling, decomposing input signals into multi-scale representations to achieve speech separation with fewer trainable parameters. TDANet \cite{li2022efficient} reduces computational cost by placing computationally intensive modules at locations with minimal temporal resolution through a downsampling-upsampling strategy. TIGER \cite{xu2024tiger} partitions frequency domain signals into multiple sub-bands for parallel processing, compressing the attention dimension from a time-frequency domain perspective.

\subsection{Causal Speech Separation}


\begin{figure*}[!t]
\centering
\includegraphics[width=1.0\linewidth]{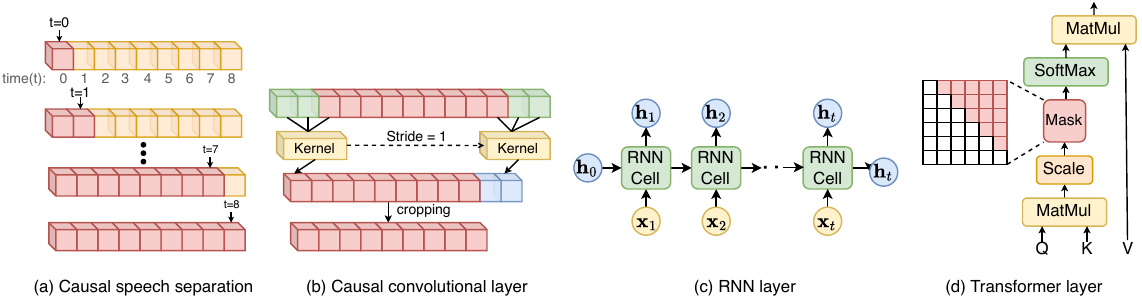}
\caption{Causal speech separation pipeline and causal network architectures. (a) Causal convolutional layers, (b) unidirectional recurrent neural networks, and Transformer models with causal masks. Causal convolutional layers apply masked convolutions along the temporal dimension to ensure that each output frame depends only on the current and previous frames. Unidirectional RNNs capture local sequential dependencies through forward recurrence. The Transformer module employs a triangular causal mask in the self-attention mechanism to block access to future frames.}
\label{fig:causal}
\vspace{-15pt}
\end{figure*}

Causal speech separation aims to address the practical requirements of speech separation technology in real-time applications \cite{healy2021causal}, as shown in Figure \ref{fig:causal}. Traditional non-causal speech separation systems typically rely on bidirectional processing mechanisms, allowing access to future frames when generating the separation output for the current time step \cite{luo2018tasnet,luo2019conv,luo2020dual,wang2023tf,zhao2024mossformer2}. While such approaches achieve superior separation performance, they inevitably introduce processing latency, significantly limiting their feasibility for deployment in applications such as hearing aids, real-time communication systems, and human-computer interaction. In contrast, causal speech separation constrains the model to utilize only current and past information, fundamentally eliminating the dependency on future frames \cite{li2022skim,li2023predictive,della2024resource}. This enables low-latency or even zero-latency speech separation, making it well-suited for real-time processing scenarios. However, this design also introduces substantial performance challenges. The absence of contextual information from future frames often results in degraded performance when handling rapidly changing speech signals or complex overlapping scenarios \cite{li2022skim}. This issue is particularly exacerbated in non-stationary noise and reverberant environments, where the difficulty of the separation task further increases. Additionally, the causality constraint limits the receptive field of the model, weakening its ability to capture long-range dependencies, which may lead to suboptimal performance in tasks requiring broader contextual information.

Causal speech separation is typically achieved by making structural modifications to key components of non-causal systems. First, bidirectional recurrent neural networks (Bi-RNNs) or bidirectional long short-term memory networks (Bi-LSTMs) need to be replaced with unidirectional architectures to ensure that information propagates only forward in the temporal sequence \cite{luo2020dual}. Prior research has demonstrated that methods such as SkiM \cite{li2022skim} and pSkiM \cite{li2023predictive} improve the performance of causal speech separation by preserving historical hidden states in LSTM layers and propagating them to subsequent layers. Additionally, in convolution-based models, causal convolution or appropriate padding strategies must be adopted to prevent information leakage \cite{luo2019conv}. For Transformer-based architectures, carefully designed masking mechanisms are required to restrict self-attention to focus exclusively on current and past information \cite{subakan2021attention}. In recent years, streaming attention mechanisms have been extensively studied and have been shown to enhance the ability of causal models to leverage past information, thereby improving separation performance~\cite{della2024resource}.

Beyond the aforementioned approaches, some studies have explored the application of state space models (SSMs) \cite{chen2023neural,li2024spmamba}, such as Mamba \cite{gu2023mamba}, in causal speech separation. These models maintain causality while effectively capturing long-range dependencies, thereby improving separation quality. Furthermore, a compromise approach allows for a limited amount of future information (e.g., a few milliseconds) to improve separation performance while maintaining low latency \cite{schroter2022low}. Collectively, these strategies drive the advancement of causal speech separation technology, ensuring that real-time requirements are met while minimizing the performance degradation imposed by causality constraints.




\subsection{Generative Approaches}

In recent years, generative methods have garnered significant attention in speech separation due to their superior generalization and data modeling capabilities compared to traditional discriminative approaches. While discriminative models directly optimize evaluation metrics—such as scale-invariant signal-to-noise ratio (SI-SNR) \cite{le2019sdr} and perceptual speech quality (PESQ) \cite{rix2001perceptual}—they often struggle to generalize to unseen conditions, especially in noisy, far-field, or reverberant environments \cite{wang2024noise}. Moreover, discriminative methods may introduce perceptually unnatural artifacts that compromise the structural integrity of speech, adversely affecting downstream tasks.

Generative models, in contrast, learn the prior distribution of data without imposing rigid parametric assumptions on the output density \cite{subakan2018generative,chen2023sepdiff}. This allows them to capture complex characteristics of speech signals—including both magnitude and phase information, which is crucial for maintaining naturalness in low signal-to-noise ratio (SNR) and reverberant conditions \cite{li2023pgss,chen2023sepdiff}. As a result, generative approaches are better equipped to produce high-quality, natural-sounding speech even in challenging acoustic scenarios.

A widely explored family of generative methods is based on Generative Adversarial Networks (GANs) \cite{goodfellow2020generative}. In GAN-based speech separation, a generator network—often implemented with convolutional neural networks (CNNs) or time-domain architectures such as Conv-TasNet \cite{luo2019conv}—synthesizes speech signals, while a discriminator network evaluates the realism of the generated outputs \cite{lakandri2024exploring}. Variants like LSGAN \cite{mao2017least} and MetricGAN \cite{fu2019metricgan} have been shown to enhance metrics such as PESQ and speech intelligibility (STOI) \cite{taal2011algorithm}, particularly under adverse SNR conditions \cite{li2018cbldnn}. The Conditional Generative Adversarial Network (Conditional GAN) framework is employed to map mixed speech signals to individual sources, often integrating permutation-invariant training (PIT) to resolve the speaker permutation problem \cite{chen2018permutation,li2021generative,fan2018svsgan,deng2020conv}. However, challenges such as mode collapse and vanishing gradients remain inherent to GAN training \cite{joseph2023cycle}.

More recently, diffusion models have emerged as a promising alternative. These models employ stochastic differential equations (SDEs) to gradually add Gaussian noise to clean speech samples in a forward process and learn a reverse process to denoise and recover the original signal \cite{chen2023sepdiff,ho2020denoising,lutati2023separate,hirano2023diffusion,scheibler2023diffusion}. Diffusion models offer a more stable training process, effectively mitigating issues like mode collapse and generating speech signals with higher naturalness and fewer artifacts \cite{wang2024noise,dong2025edsep}. Nevertheless, their standard formulation requires numerous reverse sampling steps, leading to increased computational complexity—a limitation that researchers are addressing through fast inference techniques such as Fast-GeCo \cite{wang2024noise}.

The shift from discriminative to generative paradigms in speech separation has led to improved quality and robustness, making generative approaches a promising avenue for addressing the challenges inherent in complex and noisy audio environments.

\subsection{Pre-training Methods}



In real-world cocktail party scenarios, obtaining training data with target references is particularly challenging, which has led most separation models to rely on synthetic data during training. This reliance often results in a decline in model performance when applied to real data \cite{wang2024speech,yip2024towards}. Moreover, traditional speech separation methods typically process high-dimensional waveform data directly, incurring high computational costs and further limiting their deployment on resource-constrained devices \cite{wang2024speech,yip2024towards,fazel2023cocktail}. To address these issues, researchers have proposed speech separation methods based on pre-trained models and autoencoders, which leverage the knowledge acquired from large-scale pre-training while reducing computational complexity.

Speech separation approaches based on pre-trained models have attracted extensive attention in recent years, as they effectively mitigate the scarcity of high-quality training data. Self-supervised pre-trained models, such as wav2vec 2.0 \cite{baevski2020wav2vec} and HuBERT \cite{hsu2021hubert}, acquire rich speech representations by learning from large amounts of unlabeled speech data. These representations can be transferred to the speech separation task, thereby reducing the need for task-specific labeled data \cite{huang2022investigating}. In addition, the domain gap between synthetic and real data is an important factor affecting separation performance. To this end, researchers have proposed a Domain-Invariant Pre-training (DIP) front-end, which pre-trains the feature extractor on mixed data to ensure consistent representation across real and synthetic data, thereby enhancing the generalization performance of the model \cite{wang2024speech}. In practical applications, the pre-trained model is typically employed as a feature extractor to convert the mixed audio into context-dependent embedding features that are subsequently used by the separation model to predict masks or embeddings for each speaker, or it is fine-tuned to better adapt the model to the requirements of the speech separation task \cite{huang2022investigating}. This approach not only capitalizes on the robust capabilities of pre-trained models in speech processing tasks but also reduces training costs through transfer learning.

Concurrently, autoencoder-based speech separation methods offer an efficient alternative by compressing the audio signal into a low-dimensional embedding space. Neural Audio Codecs (NAC), pre-trained on large-scale data, produce embeddings that capture features beneficial for speech processing, making separation in the compressed space feasible \cite{yip2024towards,yip2024speech}. Due to the significant temporal compression achieved by the neural codec, the sequence length in the embedding space is substantially reduced, thereby lowering the memory and computational demands of the separation model \cite{yip2024speech}.

However, methods for speech separation based on pre-trained models and autoencoders also face certain challenges. For example, autoencoders are typically trained on clean speech data and may not adequately represent overlapping speech mixtures, which limits separation performance \cite{yip2024towards,yip2024speech}. Additionally, the distortion introduced during the compression process can negatively impact separation quality, particularly in complex acoustic environments \cite{huang2022investigating,yip2024towards}.

\subsection{Target Speaker Extraction}

Target Speaker Extraction (TSE), a field closely related to speech separation, aims to isolate a specific speaker's voice from a multi-speaker mixture~\cite{zhang2025multi, he2024hierarchical, zmolikova2023neural}. To identify the target, TSE systems leverage auxiliary cues such as a pre-recorded anchor utterance~\cite{zhang2025multi, he2024hierarchical, zmolikova2023neural, he2020speakerfilter}, spatial information~\cite{ge2022spex, gu2019neural}, or visual data like lip movements~\cite{pan2021muse}. Compared to generic speech separation, TSE inherently bypasses critical issues like permutation ambiguity and speaker count dependency, making it a highly active research area with significant value for robust speech recognition and automated meeting transcription~\cite{vzmolikova2019speakerbeam, shen2025listen}.

Current TSE methods predominantly rely on audio-based cues, where an effective speaker embedding is extracted from an anchor utterance to guide the separation network~\cite{zmolikova2023neural, he2020speakerfilter, ao2024used}. Implementations vary, spanning time-domain models like the SpEx+ series~\cite{xu2020spex, ge2021multi} and frequency-domain techniques such as VoiceFilter~\cite{wang2018voicefilter} and SpeakerBeam~\cite{vzmolikova2019speakerbeam}. A central challenge lies in learning discriminative and robust speaker embeddings from limited or low-quality anchors, as simplistic single-vector representations can lead to speaker identity confusion~\cite{pandey2020cross}. To mitigate this, researchers have explored more advanced representations, including hierarchical structures~\cite{he2024hierarchical} and sparse LDA transformations~\cite{zhang2025multi}, to enhance discriminative power. While audio-only methods are convenient to deploy, their performance is critically dependent on the anchor quality.

Beyond audio-only approaches, research has expanded to multi-modal and spatial cues to handle more challenging scenarios. Audio-visual speaker extraction (AVSE) leverages visual information like lip movements, offering strong target indication in highly noisy or overlapped conditions~\cite{pan2021muse, li2022vcse, li2024audio, li2023iianet, pegg2023rtfs, pegg2023tdfnet}. However, AVSE requires high-quality, synchronized audio-visual data and incurs additional computational costs~\cite{li2023iianet}. Alternatively, methods utilizing spatial information from microphone arrays, such as L-SPEX~\cite{ge2022spex}, provide strong physical discriminability but often assume the target's location is known or necessitate an extra localization module~\cite{gu2024rezero}. These diverse strategies highlight a trade-off between performance robustness, implementation complexity, and scenario applicability.

Future TSE research holds several promising directions. The joint modeling of speaker extraction with diarization could enable systems to comprehensively answer ``who spoke what and when.'' Adopting generative frameworks, particularly conditional diffusion models, may offer enhanced robustness and generalization to unseen speakers and noise. Furthermore, expanding multi-modal fusion to include gestures or semantic cues could address extreme conditions like severe overlap or occlusion. Finally, leveraging self-supervised \cite{hu2021class} and few-shot learning will be crucial for reducing reliance on high-quality anchor data, thereby improving the practicality of TSE for diverse, real-world applications.

\subsection{Joint Modeling with Other Tasks}

Speech processing in complex auditory scenes has recently attracted significant attention.
Traditionally, speech separation and downstream tasks---such as automatic speech recognition (ASR) \cite{wu2021investigation}, speaker identification (SID) \cite{mowlaee2012joint}, and speaker diarization (SD) \cite{park2022review}---were handled in a cascaded pipeline.
This approach, however, suffers from error accumulation and mismatched optimization objectives, hindering overall performance.
Consequently, joint end-to-end modeling of speech separation with downstream tasks has become a major research trend.
These tasks prominently include ASR \cite{shi2022train, wang2015joint, berger2023mixture, raj2021integration, jiang2025speech, cornell2024one, meng2024empowering}, SD \cite{maiti2023eend, kalda2024pixit, raj2021integration, boeddeker2024ts, cornell2024one}, SID \cite{mowlaee2012joint}, speaker counting \cite{maiti2023eend}, and broader speech enhancement \cite{quan2024spatialnet}.
The core motivation is mutual benefit: separation provides cleaner inputs for downstream modules, while downstream tasks impose beneficial constraints (e.g., linguistic or speaker cues) to guide the separation process.
This synergy can even lead to unified solutions for the complex ``who spoke when and what'' problem \cite{cornell2024one, maiti2023eend}.

Existing joint training methods fall into three main categories: \textit{cascade-based}, \textit{monolithic}, and \textit{hybrid}.
Cascade-based methods \cite{shi2022train, wang2015joint} maintain modular independence while enabling global end-to-end optimization.
Monolithic approaches \cite{berger2023mixture} use a single network to bypass explicit separation, avoiding potential errors but at the cost of increased model complexity.
Hybrid methods, such as EEND-SS \cite{maiti2023eend}, TS-SEP \cite{boeddeker2024ts}, and PixIT \cite{kalda2024pixit}, aim to balance the interpretability of separation with the performance gains of joint optimization.
However, joint training introduces new challenges: balancing multi-task loss functions, managing increased computational complexity and cost, and addressing inference latency for real-time applications \cite{shi2022train, quan2024spatialnet}.

Looking forward, this research area is expected to advance in several key directions. First, foundational model adaptation methods—characterized by large models in combination with separation, adaptation/fine-tuning (e.g., soft prompts, side adapters, parameter-efficient learning)—will enable large pre-trained ASR models such as Whisper to acquire multi-speaker and target speaker recognition capabilities efficiently, with zero- or few-shot transfer to multilingual and multi-task scenarios \cite{meng2024empowering,polok2025target,guo2024sq,ma2025enhancing}. Second, joint modeling with a broader array of downstream tasks, such as speech translation, emotion recognition, and speech synthesis \cite{jiang2025speech}, may give rise to new applications. Furthermore, improving the joint processing capabilities for long-duration, continuous conversations is necessary, including the development of more efficient online streaming solutions and more robust speaker tracking \cite{cornell2024one, kalda2024pixit}. Finally, with the advancement of self-supervised and weakly supervised learning, future joint models may reduce their dependence on labeled data, thus becoming more practical and effective for real-world deployment \cite{kalda2024pixit}.

\section{Conclusions}
\label{sec:conclusion}
In this survey, we systematically survey deep neural network-based speech separation technologies, encompassing their core learning paradigms, key model architectures, mainstream evaluation metrics, datasets, and open-source tools. We not only categorize and summarize various methodologies, but also provide an in-depth analysis of the complete technical pipeline from encoder to separator to decoder. Furthermore, we conduct a comparative analysis of different learning frameworks, including supervised, self-supervised, and unsupervised approaches, with the aim of establishing a clear and comprehensive body of knowledge in this domain. Notably, we identify several existing challenges and technological gaps in current research, such as the effective handling of long-duration audio, the design of lightweight models suitable for resource-constrained devices, and the fulfillment of real-time causal processing requirements while maintaining separation performance. Based on these observations, we further delineate potential future research directions, including the exploration of more efficient generative models (e.g., diffusion models), leveraging large-scale pre-trained models and neural codecs to bridge the gap between synthetic data and real-world scenarios, as well as strengthening the joint modeling of speech separation with downstream tasks (such as speech recognition and speaker diarization). We believe this survey offers researchers and practitioners a comprehensive knowledge map and a clear technical roadmap, thereby inspiring and advancing further exploration and development in the field of speech separation across theoretical innovation, model optimization, and real-world application.


\bibliographystyle{IEEEtran}
\bibliography{refs}


 

\begin{IEEEbiography}[{\includegraphics[width=1in,height=1.25in,clip,keepaspectratio]{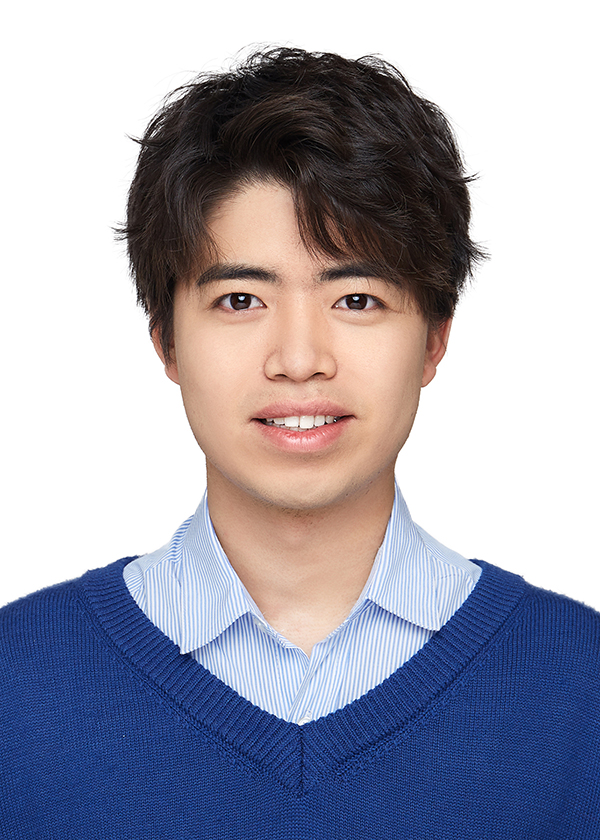}}]{Kai Li}
(Student Member, IEEE) received the B.S. degree from the Department of Computer Technology and Application, Qinghai University, Xining, China, in 2020, and the M.S. degree from the Department of Computer Science and Technology, Tsinghua University, Beijing, China, in 2024, where he is currently pursuing the Ph.D. degree under the supervision of Prof. Xiaolin Hu. His current research interests include speech/music separation, multi-modal speech separation, and audio large language models. He serves as a reviewer for several prestigious conferences and journals, including NeurIPS, ICLR, ICASSP, Interspeech, AAAI, TASLP, and TPAMI. 
\end{IEEEbiography}

\begin{IEEEbiography}[{\includegraphics[width=1in,height=1.25in,clip,keepaspectratio]{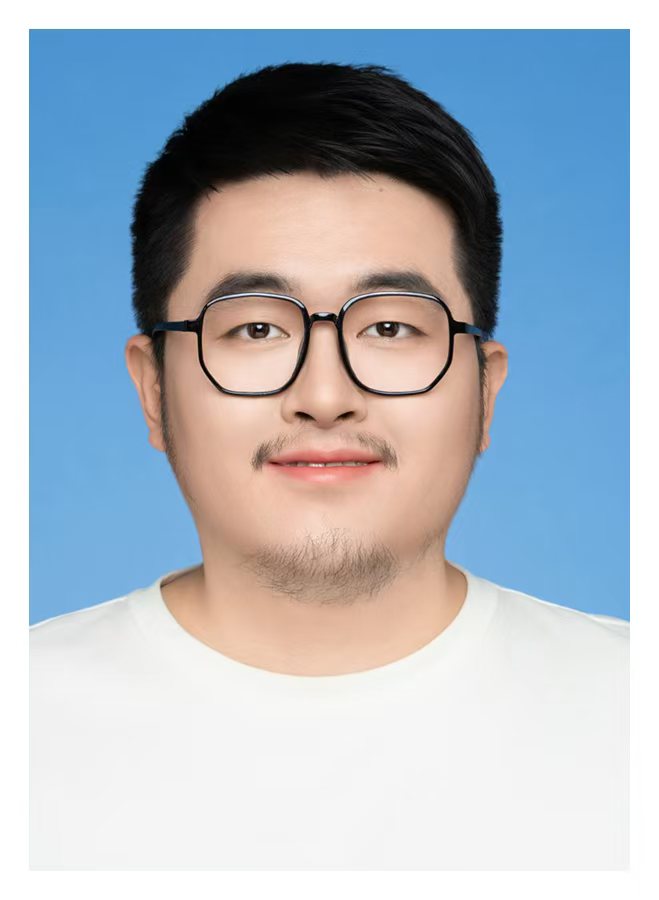}}]{Guo Chen}
with Bachelor's and Master's degrees in Computer Science and Technology from Tsinghua University, currently works at Hypergryph, where involved in the development of voice-related features in game applications. My research focused on speech technologies, including speaker separation, automatic speech recognition, and speech synthesis.
\end{IEEEbiography}

\begin{IEEEbiography}[{\includegraphics[width=1in,height=1.25in,clip,keepaspectratio]{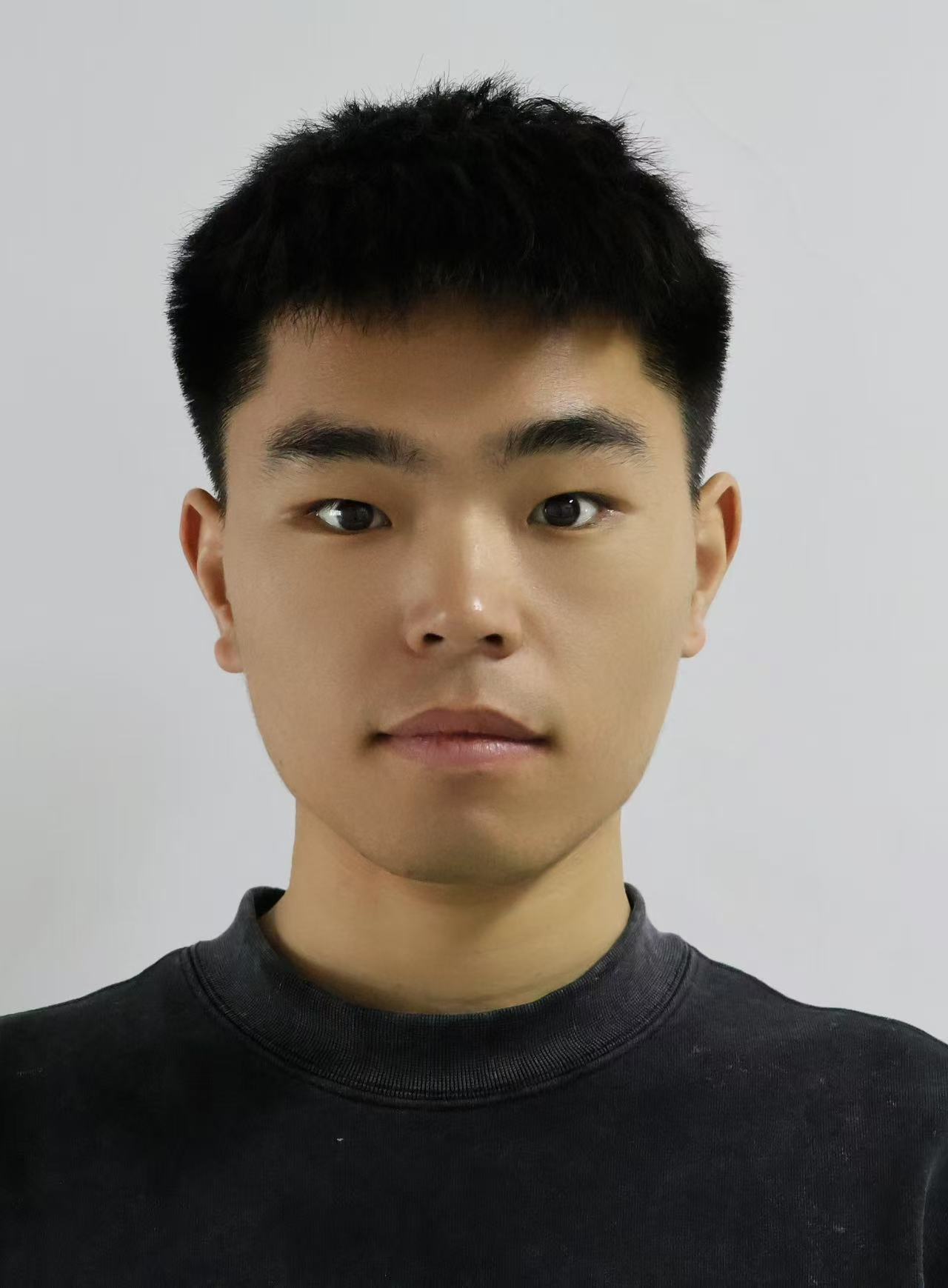}}]{Wendi Sang}
received the B.E. degree from Bengbu University, China, in 2021. He is currently pursuing the M.S. degree under the joint supervision of Qinghai University and Tsinghua University, China, with the School of Computer Technology and Application, Qinghai University. His research interests include audio processing and audio-visual learning, especially multi-modal speech separation.
\end{IEEEbiography}

\begin{IEEEbiography}[{\includegraphics[width=1in,height=1.25in,clip,keepaspectratio]{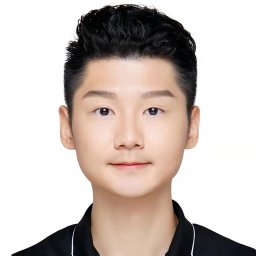}}]{Yi Luo} is currently an independent researcher. Yi obtained his Ph.D. degree from Columbia University in 2021 and then worked as a senior research scientist at Tencent AI Lab between 2021 and 2024. His main research focuses are speech and audio understanding and generation, including source separation, microphone array processing, and speech and audio generation. Yi received the 2021 IEEE Signal Processing Society Best Paper Award with his work on end-to-end speech separation.
\end{IEEEbiography}

\begin{IEEEbiography}[{\includegraphics[width=1in,height=1.25in,clip,keepaspectratio]{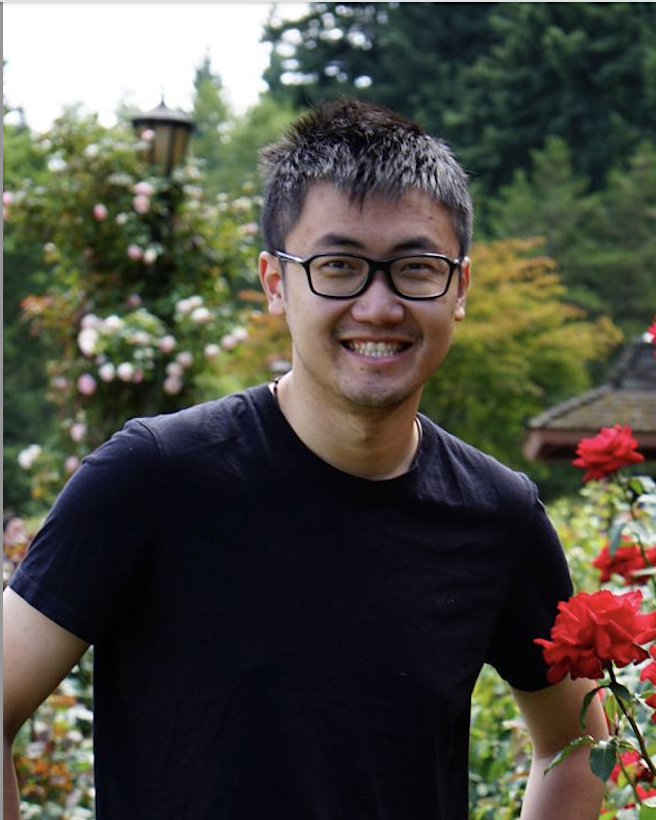}}]{Zhuo Chen}
serves as a research manager at ByteDance, leading a team focused on audio and speech generation, multi-media interaction and continual learning. Before joining ByteDance, Zhuo obtained his PhD from Columbia University in 2017 and then worked as a principal applied scientist at Microsoft. He has published over 150 research papers and patents, advancing a wide range of speech tasks including speech recognition and translation, speech separation and enhancement, speaker identification and diarization, multi-channel processing and beamforming, speech self supervised learning and speech generation.
\end{IEEEbiography}

\begin{IEEEbiography}[{\includegraphics[width=1in,height=1.25in,clip,keepaspectratio]{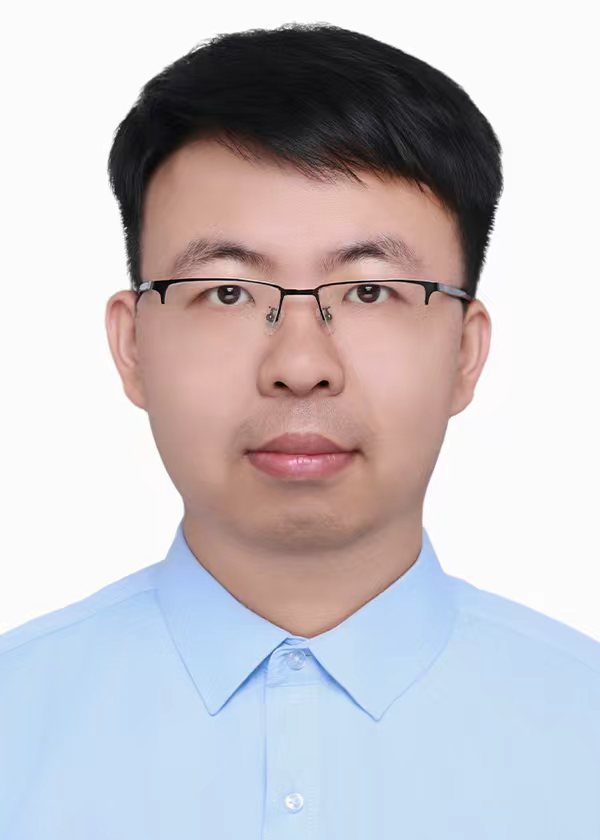}}]{Shuai Wang} (Member, IEEE) received his Ph.D. degree from Shanghai Jiao Tong University (SJTU), Shanghai, China, in 2020. He is currently an Associate Professor with Nanjing University, Suzhou, China. He was formerly a Research Scientist with the Shenzhen Research Institute of Big Data, The Chinese University of Hong Kong (Shenzhen), where he continues to hold an adjunct appointment. Prior to that, he worked at Tencent's Lightspeed \& Quantum Studios as a Senior Application Research Scientist, where he led the speech team in developing game-oriented speech technologies.  Dr. Wang's research interests include speaker modeling, target speaker separation, and speech generation. He has authored over 70 papers in premier conferences and journals in the speech processing field. He was the recipient of the IEEE Ganesh N. Ramaswamy Memorial Student Grant Award at ICASSP 2018. He was also the main contributor to the champion systems of VoxSRC 2019 and DIHARD 2019.  He is a member of ISCA, SPS and IEEE, serving as a regular reviewer for conferences and journals including ICASSP, INTERSPEECH, ASRU, TASLP and CSL. He initiated the popular ``Wespeaker'' and ``WeSep" projects, utilized by numerous research groups across academia and industry. 
\end{IEEEbiography}

\begin{IEEEbiography}[{\includegraphics[width=1in,height=1.25in,clip,keepaspectratio]{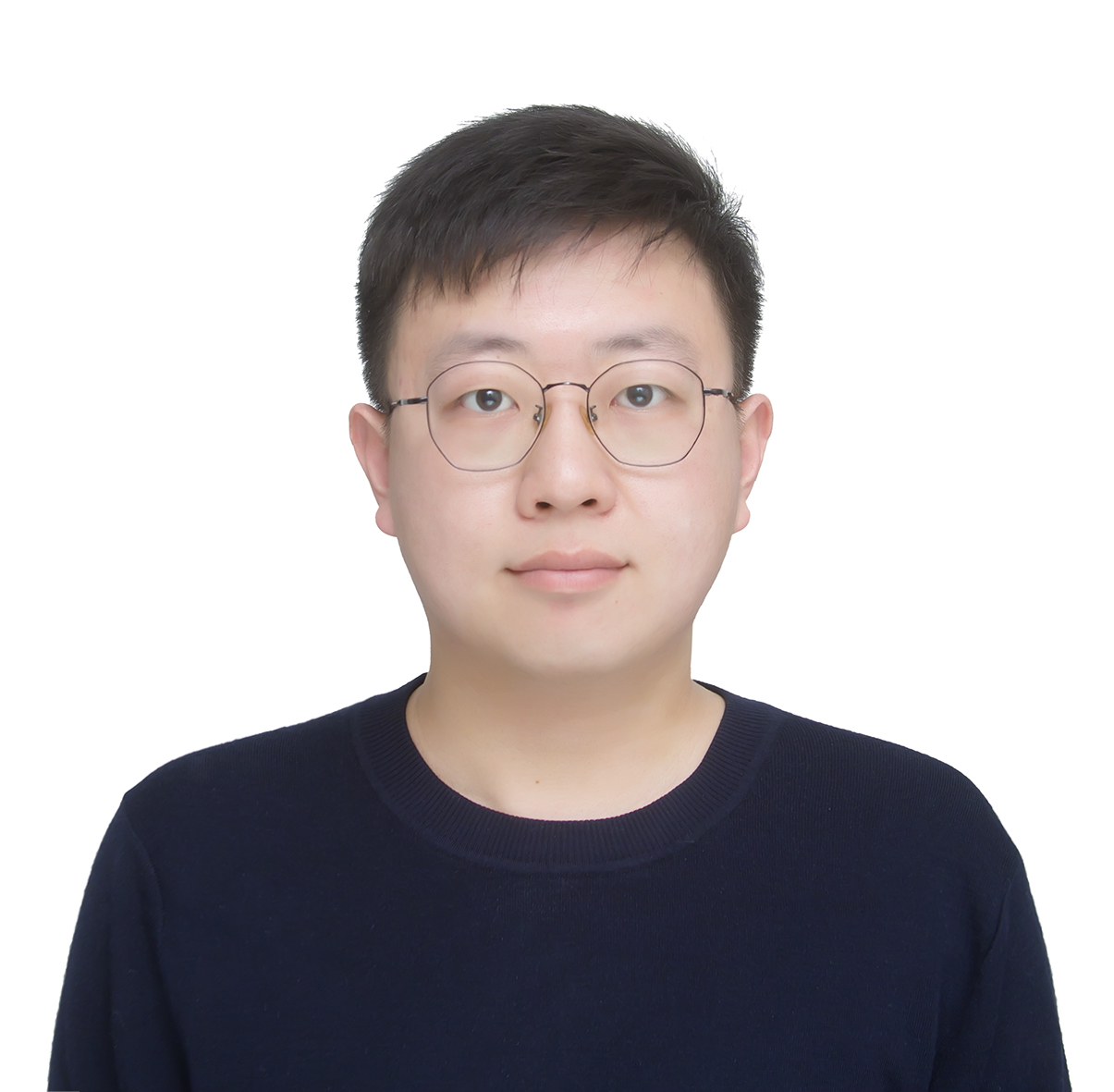}}]{Shulin He}
received the B.Eng. degree in Computer Science from Inner Mongolia University (IMU), Hohhot, China, in 2019, and the Ph.D. degree in Computer Science and technology from IMU in 2025. From June 2021 to September 2021, he was a Visiting Researcher with the Key Laboratory of Pattern Recognition, Institute of Automation, Chinese Academy of Sciences, Beijing, China. From May 2022 to May 2023, he joined the Tencent Rhino-Bird Elite Talent Program as a joint Ph.D. trainee and received its Excellent Student Award. Between October 2023 and April 2024, he was a Visiting Ph.D. Student in the Division of Emerging Interdisciplinary Areas, Hong Kong University of Science and Technology. He is currently a Postdoctoral Researcher with the SUSTech Audio Intelligence Lab, Department of Computer Science and Engineering, Southern University of Science and Technology (SUSTech), Shenzhen, China. His research interests include target speaker extraction, speech enhancement, and deep learning for speech and audio, with a focus on advancing robust speech technology in complex acoustic environments.
\end{IEEEbiography}

\begin{IEEEbiography}[{\includegraphics[width=1in,height=1.25in,clip,keepaspectratio]{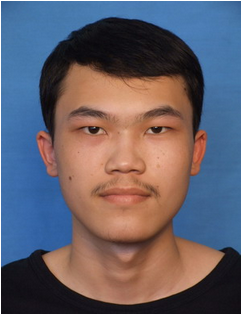}}]{Zhong-Qiu Wang}
received the B.E. degree in computer science and technology from Harbin Institute of Technology, Harbin, China, in 2013, and the M.S. and Ph.D. degrees in computer science and engineering from The Ohio State University, Columbus, OH, USA, in 2017 and 2020, respectively. He is currently a tenure-track associate professor in the Department of Computer Science and Engineering at Southern University of Science and Technology, Shenzhen, Guangdong, China. He was a Postdoctoral Research Associate with Carnegie Mellon University, Pittsburgh, PA, USA, from $2021$ to $2024$, and a visiting research scientist at Mitsubishi Electric Research Laboratories, Cambridge, MA, USA, from $2020$ to $2021$. His research interests include speech separation, robust automatic speech recognition, microphone array processing, and deep learning, aiming at solving the cocktail party problem.
\end{IEEEbiography}

\begin{IEEEbiography}[{\includegraphics[width=1in,height=1.25in,clip,keepaspectratio]{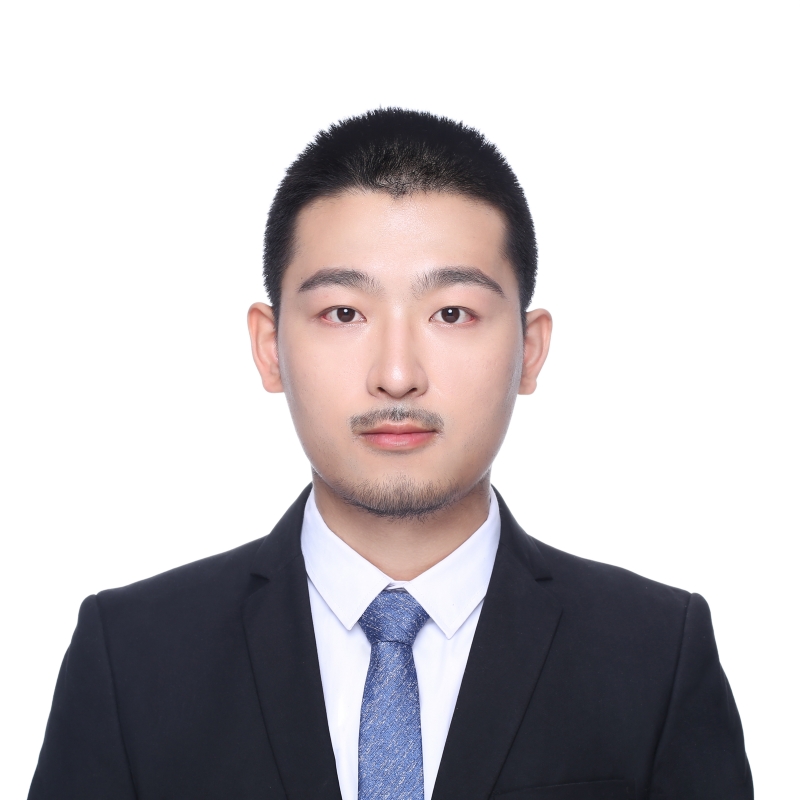}}]{Andong Li}
(Member, IEEE) received the B.S. degree in information engineering from Southeast University, Nanjing, China, in 2018, and the Ph.D. degree in signal and information processing from the Institute of Acoustics, Chinese Academy of Sciences, Beijing, China, in 2023. From 2023 to 2024, he was a Senior Researcher with Tencent AI Lab. He is currently an Associate Researcher with the Institute of Acoustics, Chinese Academy of Sciences. His research interests include speech enhancement, audio coding, array signal processing, and audio generation. He is also an active reviewer for multiple leading conferences and journals, such as INTERSPEECH, ICASSP, ICLR, ACMMM, IJCAI, AAAI, ASRU, IEEE Signal Processing Letters, Speech Communication, Neural Networks, Pattern Recognition, and IEEE/ACM Transactions on Audio, Speech, and Language Processing.
\end{IEEEbiography}

\begin{IEEEbiography}[{\includegraphics[width=1in,height=1.25in,clip,keepaspectratio]{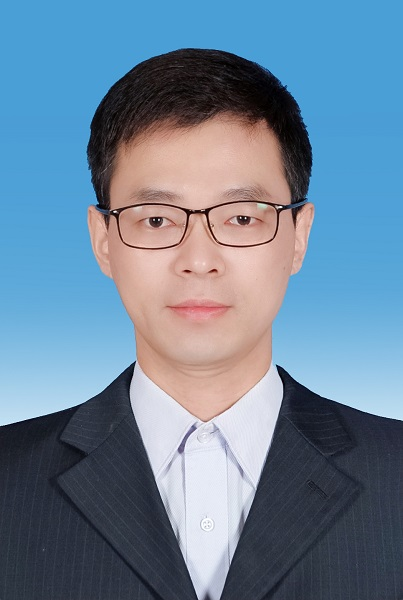}}]{Zhiyong Wu}
(Member, IEEE) received the B.S. and Ph.D. degrees in computer science and technology from Tsinghua University, Beijing, China, in 1999 and 2005, respectively. From 2005 to 2007, he was a Postdoctoral Fellow with the Department of Systems Engineering and Engineering Management, The Chinese University of Hong Kong (CUHK), Hong Kong. He then joined the Graduate School at Shenzhen (now Shenzhen International Graduate School), Tsinghua University, Shenzhen, China, and is currently a Professor. He is also a Coordinator with the Tsinghua-CUHK Joint Research Center for Media Sciences, Technologies and Systems. His research interests include intelligent speech interaction, more specially, speech processing, audiovisual bimodal modeling, text-to-audio-visual-speech synthesis, and natural language understanding and generation. He is a Member of International Speech Communication Association (ISCA) and China Computer Federation (CCF).
\end{IEEEbiography}

\begin{IEEEbiography}[{\includegraphics[width=1in,height=1.25in,clip,keepaspectratio]{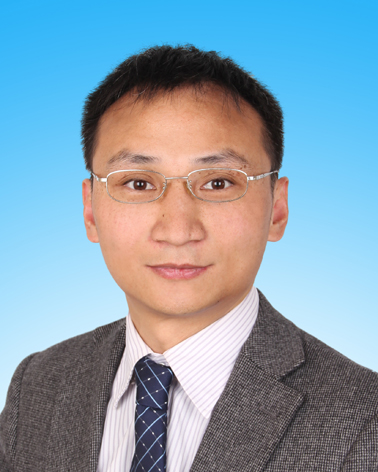}}]{Xiaolin Hu}
(S’01-M’08-SM’13) received B.E. and M.E. degrees in automotive engineering from the Wuhan University of Technology, Wuhan, China, in 2001 and 2004, respectively, and a Ph.D. degree in automation and computer-aided engineering from the Chinese University of Hong Kong, Hong Kong, in 2007. He is currently an Associate Professor at the Department of Computer Science and Technology, Tsinghua University, Beijing, China. His current research interests include deep learning and computational neuroscience. At present, he is an Associate Editor of the IEEE Transactions on Pattern Analysis and Machine Intelligence and Cognitive Neurodynamics. Previously he was an Associate Editor of the IEEE Transactions on Image Processing, IEEE Transactions on Neural Networks and Learning Systems.
\end{IEEEbiography}



\vfill

\end{document}